\documentclass[%
preprint,
 amsmath,amssymb,
 aps,
]{revtex4-2}

\usepackage{microtype}
\usepackage{comment}
\usepackage{natbib}
\usepackage{graphicx}
\usepackage{caption}
\usepackage{ragged2e}
\DeclareCaptionJustification{myjustified}{\justifying}
\usepackage{dcolumn}
\usepackage{bm}
\usepackage{empheq}
\usepackage{makecell}
\usepackage{adjustbox}
\usepackage{booktabs}
\usepackage{tabularx}
\usepackage{xcolor}
\usepackage{array}
\usepackage{rotating}
\usepackage{placeins}
\usepackage{multirow}
\usepackage[hidelinks]{hyperref}

\makeatletter
\renewcommand{\bibsection}{%
  \section*{References}%
}
\makeatother

\begin{document}


\title{Inverse heterodyne effect in\\
Bimodal Kelvin Probe Force Microscopy}

\author{
Hugo Valloire$^{(1)}$, Sylvain Clair$^{(1)}$, Christian Loppacher$^{(1)}$ and Laurent Nony$^{(1),\dagger}$ \\
Benjamin Gr\'evin$^{(2),\ddagger}$\\
{\scriptsize $^\dagger$\texttt{laurent.nony@im2np.fr}}\\
{\scriptsize $^\ddagger$\texttt{benjamin.grevin@neel.cnrs.fr}}
}

\affiliation{%
\vspace{0.5em}%
\begin{tabular}{c}
$^{(1)}$ Aix Marseille Université, CNRS, Univ Toulon, IM2NP, Marseille, France \\[0.25em]
$^{(2)}$ Institut Néel, UPR 2940 CNRS, Grenoble, France
\end{tabular}%
\vspace{0.5em}%
}

\date{\today}

\begin{abstract}
Heterodyne Kelvin probe force microscopy (He-KPFM) enables high-sensitivity electrostatic measurements by converting a bias-modulated interaction into a resonant response at a higher cantilever eigenmode. While the ``direct'' heterodyne actuation of the second eigenmode is well established, the dynamical back-action of this heterodyne-driven motion on the fundamental eigenmode has remained largely unexplored, particularly in open-loop operation where the second mode is excited to a finite amplitude. Here, an inverse heterodyne effect is demonstrated: a force component generated by heterodyne frequency conversion acts back on the first eigenmode and produces measurable inter-mode energy exchange. The analysis combines the bimodal virial and power-balance framework with a non-truncated description of the dynamics of the tip-surface capacitance gradient developed and validated in a companion manuscript submitted concurrently to the same journal, which provides the full derivation and regimes of validity of the effective capacitance-gradient coefficients used here. Building on this foundation, closed-form expressions are derived linking inverse heterodyne coupling to the experimentally accessible observables of non-contact AFM open-loop amplitude-modulated He-KPFM. The theory predicts that inverse heterodyne coupling manifests predominantly in the dissipation channel, with a sharply resonant dependence on the demodulation frequency near the second-eigenmode resonance, while its conservative contribution to the frequency shift is comparatively weaker under typical conditions. Ultrahigh-vacuum experiments validate these predictions and isolate the inverse heterodyne signature through frequency- and voltage-dependent measurements.
\end{abstract}

\maketitle

\clearpage

\section{Introduction}\label{sec:introduction}

Kelvin probe force microscopy (KPFM) has become a workhorse technique for mapping surface electrostatic potentials and charge distributions with nanoscale resolution, with applications ranging from semiconductor physics to molecular and surface science~\cite{Nonnenmacher1991a,Melitz2011a,Sadewasser2012a,Sadewasser2018a}. In bias-modulated implementations, an AC voltage is applied between tip and sample, generating electrostatic force components whose demodulated response provides access to the contact potential difference (CPD)~\cite{Kelvin1898a,Zisman1932a} and capacitance-related quantities~\cite{Magonov2011,Borgani2014a,Collins2015_3DKPFM,Rohrbeck2025_MFHEFM}. Over the years, KPFM has evolved into a broad family of detection schemes, including amplitude-modulated (AM) and frequency-modulated (FM) variants~\cite{Kikukawa1996a,kitamura98a,zerweck05a,Kilpatrick2018a}.

A major advance in sensitivity and robustness has been achieved with heterodyne approaches, which exploit electromechanical frequency conversion between a mechanically driven cantilever oscillation and an electrically modulated bias~\cite{Sugawara2012a,Borgani2014a}. In amplitude-modulated heterodyne KPFM (AM-He-KPFM), the bias-modulation frequency is chosen such that one of the electrostatic intermodulation products falls on a higher cantilever eigenmode, enabling resonant detection of a signal proportional to the electrostatic force gradient~\cite{Sugawara2012a}. Heterodyne concepts also underpin more recent developments, including dual heterodyne (DHe) and photo-induced schemes that probe intermodulation products generated under external pumping~\cite{Borgani2014a,Borgani2019a,Grevin2023a}.

Although open-loop AM-He-KPFM provides a practical route to heterodyne KPFM measurements, its dynamical interpretation remains incomplete. Existing descriptions commonly treat the higher eigenmode as a passive readout channel driven by an electrostatic force component generated by the fundamental-mode motion and the applied bias~\cite{Sugawara2012a,Borgani2014a}. This viewpoint tends to overlook the fact that, in open-loop operation, the higher eigenmode is excited to a finite amplitude and thereby contributes to the tip-surface distance modulation. In other contexts of multifrequency atomic force microscopy (AFM), it is well established that nonlinear interactions can transfer energy between eigenmodes and produce measurable back-action~\cite{Lozano2009a,Herruzo2012a,Kiracofe2013a,An2014a}. Whether, and how, an analogous inter-mode coupling arises from heterodyne electrostatic actuation has not been clarified.

It is shown here that open-loop AM-He-KPFM intrinsically supports an inverse heterodyne effect, in which the electrostatically sustained second eigenmode feeds back onto the conservative and dissipative dynamics of the first eigenmode. This mechanism implies that heterodyne excitation modifies not only the heterodyne-detected response at the higher eigenmode, but also the fundamental-mode observables used in non-contact AFM (nc-AFM), namely frequency shift and energy dissipation. Because dissipation is directly tied to the quadrature component of the interaction force in the virial/power-balance framework~\cite{giessibl00a,Lozano2009a}, inverse heterodyne coupling is expected to leave a signature in the dissipation channel.

Dissipative implementations of KPFM have shown that electrostatic dissipation can itself be exploited as an efficient measurement channel, with sensitivity to either the electrostatic force or its gradient~\cite{Miyahara2015a,Miyahara2017a,Sadewasser2018a}. These approaches demonstrated that dissipation may provide a fast and practically convenient route to Kelvin probe measurements. In the present work, however, dissipation is not introduced as an alternative readout channel. Instead, it is shown that, in open-loop AM-He-KPFM, the dissipation of the first eigenmode can contain the signature of an inverse heterodyne back-action generated by the second eigenmode. The objective of the present work is therefore not to develop another dissipative KPFM modality, but rather to establish the dynamical origin and experimental consequences of inter-mode energy exchange in heterodyne bimodal operation.

To establish this effect on firm ground, the electrostatic force components acting at the two eigenmode frequencies under heterodyne biasing are derived, separating (i) the well-known direct heterodyne actuation of the second eigenmode from (ii) an inverse heterodyne contribution acting back on the first eigenmode. The treatment relies on a compact representation of the tip-surface capacitance gradient (CG) dynamics under bimodal motion; the full non-truncated derivation, truncation regimes, and numerical validation of the corresponding effective coefficients are provided in the companion manuscript submitted concurrently to the same journal~\cite{ValloireSubmittedCG}. Incorporating the resulting force expressions into the bimodal virial and power-balance equations yields closed-form predictions for the experimentally accessible observables in nc-AFM open-loop AM-He-KPFM, namely the frequency shift and dissipated energy of the first eigenmode, and the amplitude and phase of the second eigenmode.

These predictions are experimentally validated in ultrahigh vacuum (UHV) nc-AFM, and inverse heterodyne coupling is shown to produce a sharply resonant signature when sweeping the demodulation frequency around the second eigenmode resonance. The measurements further show that the inverse heterodyne effect predominantly manifests in the dissipation channel, while its conservative contribution to the frequency shift remains comparatively weaker under typical conditions. Beyond open-loop AM-He-KPFM, the results of the present study apply to heterodyne electrostatic force microscopy more broadly and provide general insight into energy exchange and back-action in driven multimode nanomechanical systems involving frequency conversion.

In addition to predicting the behavior of the observables in open-loop experiments, the present formalism also provides new insights for a more accurate analytical treatment of He-KPFM variants, including dual-heterodyne KPFM (DHe-KPFM)~\cite{Grevin2023a} and heterodyne photo-induced force microscopy (He-PiFM)~\cite{Jahng2016a,Yamanishi2017a,Shcherbakov2025a}.

The manuscript is organized as follows. Sec.~\ref{sec:minimal_framework} introduces the minimal bimodal framework and the CG toolbox used throughout. Sec.~\ref{sec:electrostatic_force_components_bimodal_dynamics} derives the electrostatic force components responsible for direct and inverse heterodyne coupling. Sec.~\ref{sec:inter_mode_energy_exchange_coupled_observables}
 connects these force components to the coupled observables through virial and power balance. Sec.~\ref{sec:experimental_signatures_inverse_heterodyne_back_action}
 presents the experimental validation and the discriminating signatures of inverse heterodyne back-action. Sec.~\ref{sec:discussion_generalization} discusses implications and generalizations to advanced heterodyne schemes. Lastly, a supplementary information file (SI) is also available, providing the details of the analytical calculations and numerical support.

\clearpage

\section{Minimal framework}\label{sec:minimal_framework}

\subsection{Bimodal AFM essentials}\label{sec:bimodal_afm_essentials}

The present analysis is based on the fundamentals of bimodal AFM~\cite{Rodriguez2004a,Lozano2008a,Lozano2009a}, the basic elements of which are only briefly reviewed here. The cantilever deflection is usually modeled as a superposition of two independent point-mass 1D oscillators (simple harmonic oscillator (SHO) model) representing the first and second cantilever eigenmodes. These modes correspond to the dominant mechanical resonances of the cantilever and are described, within the SHO approximation, by their respective transfer functions. For each eigenmode ($i=1,2$), this transfer function is written as:

\begin{subequations}\label{eq:transfer_function}
\begin{equation}
\hat{G}_i(\omega)=\frac{1}{k_i}\frac{1}{1-\frac{\omega^2}{\omega_{i,0}^2}+j\frac{\omega}{\omega_{i,0}Q_i}}
=\left|\hat{G}_i(\omega)\right|e^{j\Phi_{\hat{G}_i}(\omega)}.
\label{eq:transfer_function_a}
\end{equation}

Its modulus and phase are given by:

\begin{empheq}[left=\empheqlbrace]{align}
\left|\hat{G}_i(\omega)\right|
&= \frac{1}{k_i}
\frac{1}{
\sqrt{
\left(1-\frac{\omega^2}{\omega_{i,0}^2}\right)^2
+
\left(\frac{\omega}{\omega_{i,0}Q_i}\right)^2
}
},
\label{eq:transfer_function_b}
\\
\Phi_{\hat{G}_i}(\omega)
&=
-\operatorname{atan2}\!\left(
\frac{\omega}{\omega_{i,0}Q_i},
1-\frac{\omega^2}{\omega_{i,0}^2}
\right),
\label{eq:transfer_function_c}
\end{empheq}
\end{subequations}

\noindent where $\omega_{i,0}=2\pi f_{i,0}$, $Q_i$, and $k_i$ are, respectively, the free angular resonance frequency, quality factor, and modal stiffness associated with the $i$th eigenmode. The function $\operatorname{atan2}(y,x)$ denotes the two-argument arctangent, which returns the principal phase angle associated with the point $(x,y)$, with the quadrant determined by the signs of $x$ and $y$. At resonance, $\omega=\omega_{i,0}$, the modulus and the phase of the transfer function are given by $\left|\hat{G}_i(\omega_{i,0})\right|=Q_i/k_i$ and $\Phi_{\hat{G}_i}(\omega_{i,0})=-\pi/2$, indicating a maximum amplification governed by the quality-factor-to-stiffness ratio and a phase lag between the cantilever deflection and the driving force.

The vertical motion $z_i(t)$ of each eigenmode ($i=1,2$) is then described by the usual second-order differential equation:

\begin{subequations}\label{eq:sho_motion}
\begin{empheq}[left=\empheqlbrace]{align}
\ddot{z}_i(t)+\frac{\omega_{i,0}}{Q_i}\dot{z}_i(t)+\omega_{i,0}^2 z_i(t)
&= \frac{F_{\mathrm{tot}}(z(t),t)}{m_i},
\label{eq:sho_motion_a}
\\
\omega_{i,0}^2
&= \frac{k_i}{m_i},
\label{eq:sho_motion_b}
\\
F_{\mathrm{tot}}(z(t),t)
&= F_d(t)+F_{\mathrm{int}}(z(t))+F_{\mathrm{el}}(z(t),t),
\label{eq:sho_motion_c}
\end{empheq}
\end{subequations}

\noindent where $m_i$ is the effective mass of the $i$th eigenmode ($i=1,2$). For rectangular beam cantilevers, the first two eigenmodes have identical effective masses, equal to one quarter of the total cantilever mass~\cite{melcher07a,Lozano2009a}. In the following, the assumption $m_1=m_2=m_{\mathrm{eff}}$ is therefore made.

In Eq.~\eqref{eq:sho_motion_a}, $F_{\mathrm{tot}}(z(t),t)$ denotes the total force exerted on the cantilever that drives its motion. It includes the external mechanical drive $F_d(t)$, the tip-surface interaction force $F_{\mathrm{int}}(z(t))$, and the electrostatic force $F_{\mathrm{el}}(z(t),t)$, see Eq.~\eqref{eq:sho_motion_c}.

\begin{figure}[htbp]
    \centering
    \includegraphics[width=\textwidth]{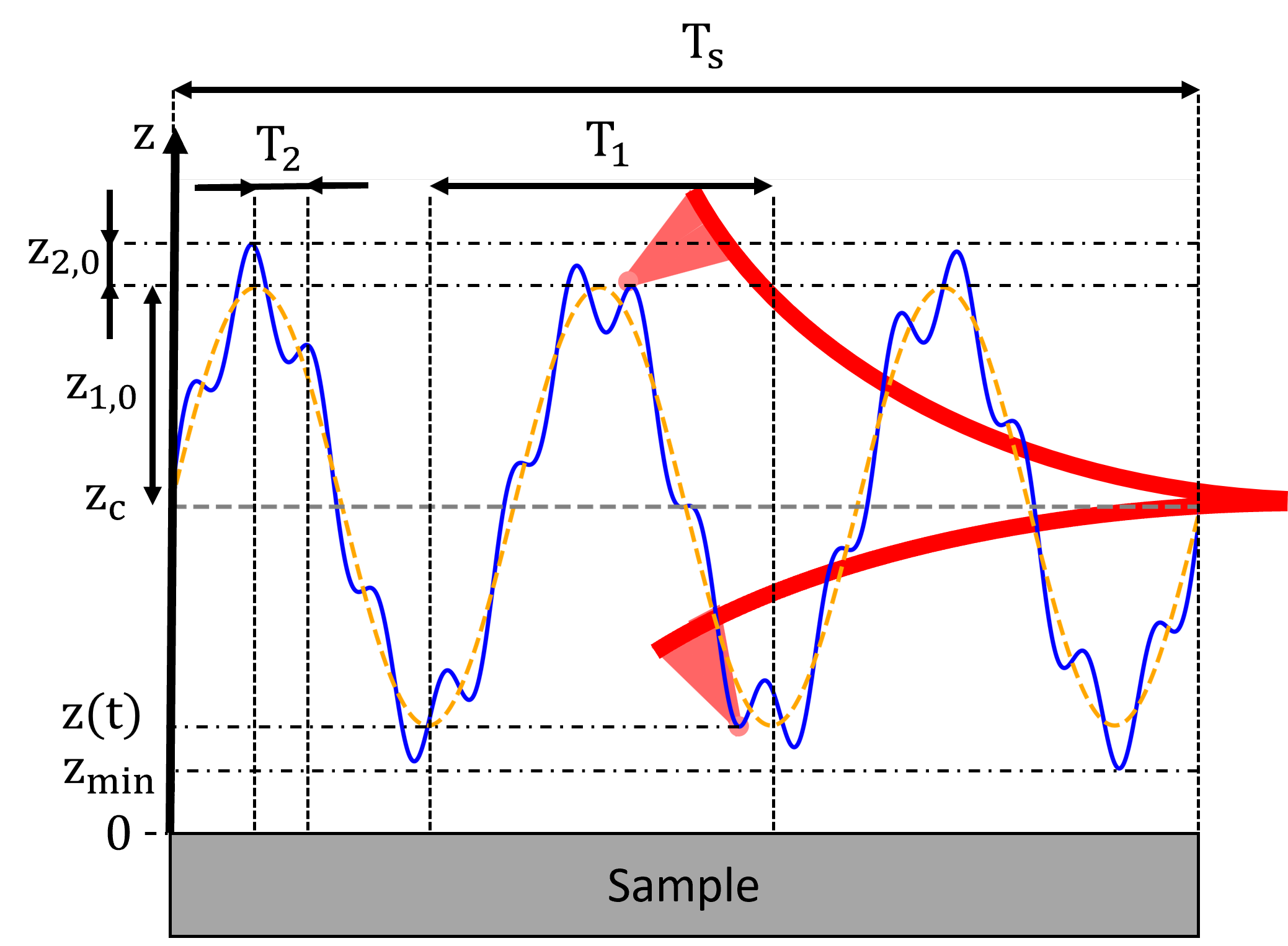}
    \caption{Geometry of the problem. Schematic representation of the cantilever oscillatory motion above the sample surface in the bimodal regime. The blue solid curve represents the instantaneous tip trajectory, $z(t)$, resulting from the superposition of the oscillations of the first and second eigenmodes. The yellow dashed curve illustrates the oscillatory component associated with the first eigenmode. The oscillation amplitudes of the first and second eigenmodes are denoted $z_{1,0}$ and $z_{2,0}$, respectively. The gray dashed horizontal line indicates the average tip--surface distance, $z_c$. The quantity $z_{\min}$ denotes the minimum tip--surface distance. The corresponding oscillation periods are $T_1=2\pi/\omega_{1,0}$ and $T_2=2\pi/\omega_{2,0}$, while $T_s$ denotes the super-period of the bimodal motion.}
    \label{fig:geometry_problem}
\end{figure}

The geometry of the problem is shown in Fig.~\ref{fig:geometry_problem}. When restricting the mechanical deflection of the cantilever to its first two eigenmodes, the total deflection can be approximated as a superposition of the oscillations induced by each eigenmode, such that the instantaneous tip-surface distance $z(t)$ is defined as:

\begin{subequations}\label{eq:tip_distance}
\begin{empheq}[left=\empheqlbrace]{align}
z(t) &= z_c + \sum_{i=1}^{2} z_i(t),
\label{eq:tip_distance_a}
\\
z_c &= z_{\min} + \sum_{i=1}^{2} z_{i,0},
\label{eq:tip_distance_b}
\\
z_i(t) &= z_{i,0}\cos(\omega_i t + \Phi_i).
\label{eq:tip_distance_c}
\end{empheq}
\end{subequations}

In Eq.~\eqref{eq:tip_distance_a}, $z_c$ denotes the average tip-surface distance. The minimum tip-surface distance during the oscillation cycle of the cantilever is denoted by $z_{\min}$. Each eigenmode oscillation is characterized by its amplitude, angular frequency, and phase ($z_{i,0}$, $\omega_i=2\pi f_i$, and $\Phi_i$, respectively), which are among the physical observables of the problem. Note, however, that the tip-surface distance entering the expression of the total force is $z(t)$ and not $z_i(t)$ (see Eq.~\eqref{eq:sho_motion_c}).

Eq.~\eqref{eq:tip_distance} is sufficient to describe the dynamics of the cantilever in bimodal AFM when no modulation of the electrostatic force occurs. However, as discussed in the introduction, in KPFM the modulation of the electrostatic force induces additional oscillatory components in the cantilever dynamics. Thus, Eq.~\eqref{eq:tip_distance} provides only an approximate description of the actual motion. Nevertheless, this representation is assumed to remain valid, since the cantilever eigenmodes constitute the dominant contributions to its deflection.

To determine the AFM observables induced by the force components, Eq.~\eqref{eq:sho_motion} applied to each eigenmode ($i=1,2$) can be solved, under the harmonic assumption (Eq.~\eqref{eq:tip_distance_c}), either by invoking the virial theorem and energy conservation~\cite{Lozano2009a}, or by projecting onto the orthogonal basis $(\cos(\omega_i t+\Phi_i),\ \sin(\omega_i t+\Phi_i))$. Both procedures yield identical results in the form of a pair of coupled equations for each eigenmode (see SI, Sec.~SI-I):

\begin{subequations}\label{eq:coupled_observables}
\begin{empheq}[left=\empheqlbrace]{align}
\frac{\Delta\omega_i(z_c)}{\omega_{i,0}}
&= -\frac{1}{z_{i,0}k_iT}
\int_0^T F_{\mathrm{tot}}(t,z_c)\cos(\omega_i t+\Phi_i)\,dt,
\label{eq:coupled_observables_a}
\\
\frac{z_{i,0}\omega_i(z_c)}{2Q_i\omega_{i,0}}
&= -\frac{1}{k_iT}
\int_0^T F_{\mathrm{tot}}(t,z_c)\sin(\omega_i t+\Phi_i)\,dt.
\label{eq:coupled_observables_b}
\end{empheq}
\end{subequations}

In the following, Eq.~\eqref{eq:coupled_observables_a} is referred to as the virial equation, whereas Eq.~\eqref{eq:coupled_observables_b} is referred to as the energy or power equation. Here, the frequency shift $\Delta\omega_i(z_c)=\omega_i(z_c)-\omega_{i,0}$ of the eigenmode resonance frequency is introduced, together with its dependence on the average tip-surface distance $z_c$. These equations are formally identical to those derived in the monomodal case~\cite{giessibl00a}. However, in the bimodal case, they are valid provided that the integration is performed over a sufficiently long time interval $T$ that accounts for the oscillation periods of the first and second eigenmodes ($T_1$ and $T_2$), as well as the bias modulation period $T_{\mathrm{mod}}=1/f_{\mathrm{mod}}=2\pi/\omega_{\mathrm{mod}}$ that modulates the tip-surface electrostatic force.

A natural period to consider for the bimodal dynamics of the cantilever is the ``super-period'', $T_s$, which establishes a commensurability relationship between the eigenmode frequencies~\cite{Lozano2009a,Kawai2009a,Lai2015a} (see also Sec.~SI-I):

\begin{equation}\label{eq:super_period}
T_s
=
p_1T_1
=
\frac{2\pi p_1}{\omega_1}
=
p_2T_2
=
\frac{2\pi p_2}{\omega_2},
\qquad
p_1,p_2\in\mathbb{N}^{\ast},
\qquad
\gcd\left(p_1,p_2\right)=1.
\end{equation}

The additional periodicity introduced by the bias modulation can now be accounted for. The long time interval $T$ ultimately considered to solve Eq.~\eqref{eq:coupled_observables_a} and Eq.~\eqref{eq:coupled_observables_b} is the ``super-super-period'', $T_{ss}$, built consistently from $T_s$ and $T_{\mathrm{mod}}$, namely:

\begin{equation}\label{eq:super_super_period}
T=T_{ss}=q_1T_s=q_2T_{\mathrm{mod}};
\qquad (q_1,q_2)\in\left(\mathbb{N}^{\ast}\right)^2.
\end{equation}

Eq.~\eqref{eq:coupled_observables_a} and Eq.~\eqref{eq:coupled_observables_b} are general expressions that can be used to calculate the oscillation amplitudes, phases, and/or frequency shifts of each eigenmode, depending on the chosen operating mode. These equations link the observables to the virial, corresponding to processes in phase with the deflection and mainly of conservative nature (Eq.~\eqref{eq:coupled_observables_a}), and to the power, or equivalently the energy supplied by the external force to the cantilever eigenmode over the considered time interval, corresponding to processes in phase quadrature with the deflection and mainly of dissipative nature (Eq.~\eqref{eq:coupled_observables_b}).

Based on the above elements, the problem can be solved once the spectral components of the total force acting on the cantilever eigenmodes at $\omega_1$ and $\omega_2$ are known. In the remainder of this paper, these components are determined, including those imposed by the modulated electrostatic force at play in KPFM.

\subsection{Electrostatic force}\label{sec:electrostatic_force}

In electrostatic force microscopy and KPFM, the attractive electrostatic force between the tip and sample is generally expressed as follows~\cite{Nonnenmacher1991a}:

\begin{subequations}\label{eq:electrostatic_force}
\begin{empheq}[left=\empheqlbrace]{align}
F_{\mathrm{el}}(z(t),t)
&= \frac{1}{2}\frac{dC(z)}{dz}\left[V_{\mathrm{DC}}-V_{\mathrm{cpd}}+V_{\mathrm{mod}}(t)\right]^2,
\label{eq:electrostatic_force_a}
\\
V_{\mathrm{mod}}(t)
&= U_{\mathrm{mod}}\cos(\omega_{\mathrm{mod}} t+\Phi_{\mathrm{mod}}).
\label{eq:electrostatic_force_b}
\end{empheq}
\end{subequations}

$V_{\mathrm{DC}}$, $V_{\mathrm{mod}}(t)$, and $V_{\mathrm{cpd}}$ respectively denote the DC bias (i.e.\ the compensation bias used in closed-loop KPFM to nullify the CPD), the AC-modulated bias, and the CPD between the tip and the surface. In Eq.~\eqref{eq:electrostatic_force_a}, the term $(V_{\mathrm{DC}}-V_{\mathrm{cpd}})$ assumes that the bias is applied to the sample (tip grounded). It must be recast as $(V_{\mathrm{DC}}+V_{\mathrm{cpd}})$ in the opposite configuration. The modulated AC bias is characterized by the modulation depth $U_{\mathrm{mod}}$, angular frequency $\omega_{\mathrm{mod}}$, and phase lag $\Phi_{\mathrm{mod}}$ (arbitrarily defined with respect to the eigenmode oscillation phases).

$C(z)$ is the tip-surface capacitance, defined with respect to the instantaneous tip-surface distance $z(t)$ introduced previously (see Eq.~\eqref{eq:tip_distance}). In the following, to simplify the notation, the CG is denoted by $C^{(1)}$, and its implicit time dependence arising from that of $z(t)$ is considered, such that $dC(z)/dz \rightarrow C^{(1)}(z) \rightarrow C^{(1)}(t,z_c)$. So far, this description has mostly relied on a Taylor series expansion of the CG truncated at first order and restricted to the monomodal dynamics of the cantilever~\cite{Bonnell2012a,Borgani2014a,Axt2018a,Garrett2018a,Garrett2019a}. Although conceptually useful, this framework remains insufficient for the present work. A detailed non-truncated derivation of $C^{(1)}(t,z_c)$ in the bimodal context is provided in the companion manuscript~\cite{ValloireSubmittedCG}.

The electrostatic force is commonly expanded into three terms corresponding to its DC, $\omega_{\mathrm{mod}}$, and $2\omega_{\mathrm{mod}}$ spectral components, namely the fundamental and second-harmonic frequencies of the AC-modulated bias:

\begin{subequations}\label{eq:electrostatic_components}
\begin{empheq}[left=\empheqlbrace]{align}
F_{\mathrm{el}}^{\mathrm{DC}}(t,z_c)
&= \frac{1}{2}C^{(1)}(t,z_c)\left[(V_{\mathrm{DC}}-V_{\mathrm{cpd}})^2+\frac{U_{\mathrm{mod}}^2}{2}\right],
\label{eq:electrostatic_components_a}
\\
F_{\mathrm{el}}^{\omega_{\mathrm{mod}}}(t,z_c)
&= C^{(1)}(t,z_c)(V_{\mathrm{DC}}-V_{\mathrm{cpd}})U_{\mathrm{mod}}
\cos(\omega_{\mathrm{mod}} t+\Phi_{\mathrm{mod}}),
\label{eq:electrostatic_components_b}
\\
F_{\mathrm{el}}^{2\omega_{\mathrm{mod}}}(t,z_c)
&= \frac{1}{4}C^{(1)}(t,z_c)U_{\mathrm{mod}}^2
\cos(2\omega_{\mathrm{mod}} t+2\Phi_{\mathrm{mod}}).
\label{eq:electrostatic_components_c}
\end{empheq}
\end{subequations}

\subsection{Capacitance gradient}\label{sec:capacitance_gradient}

As stated above, the electrostatic force depends on the CG $C^{(1)}(t,z_c)\equiv dC(z)/dz$ evaluated along the bimodal trajectory $z(t)=z_c+z_1(t)+z_2(t)$. In the present work, only the components of $C^{(1)}(t,z_c)$ that contribute at DC, $\omega_1$, and $\omega_2$ are required, and they are written in the compact form:

\begin{subequations}\label{eq:CG_expansion}
\begin{align}
C^{(1)}(t,z_c)
&=K_0(z_c)+K_1(z_c)z_1(t)+K_2(z_c)z_2(t)+\ldots
\notag\\
&=K_0(z_c)+z_{1,0}K_1(z_c)\cos(\omega_1 t+\Phi_1)
+z_{2,0}K_2(z_c)\cos(\omega_2 t+\Phi_2)+\ldots
\label{eq:CG_expansion_a}
\end{align}

The effective coefficients $K_i(z_c)$ ($i=0,1,2$) collect all higher-order derivatives of the CG with respect to $z_c$ and capture the nontrivial dependence on the oscillation amplitudes and the average tip-surface distance. Retaining the leading contribution and the first grouped correction yields the expressions in the first-order truncation regime (FOTR), derived in the companion manuscript~\cite{ValloireSubmittedCG}:

\begin{empheq}[left=\empheqlbrace]{align}
K_0(z_c)
&\approx
K_0^{\mathrm{FOTR}}(z_c)
=
C^{(1)}(z_c)
+
\frac{z_{1,0}^{2}+z_{2,0}^{2}}{4}
C^{(3)}(z_c),
\label{eq:CG_expansion_b}
\\
K_1(z_c)
&\approx
K_1^{\mathrm{FOTR}}(z_c)
=
C^{(2)}(z_c)
+
\left(
\frac{z_{1,0}^{2}}{8}
+
\frac{z_{2,0}^{2}}{4}
\right)
C^{(4)}(z_c),
\label{eq:CG_expansion_c}
\\
K_2(z_c)
&\approx
K_2^{\mathrm{FOTR}}(z_c)
=
C^{(2)}(z_c)
+
\left(
\frac{z_{1,0}^{2}}{4}
+
\frac{z_{2,0}^{2}}{8}
\right)
C^{(4)}(z_c).
\label{eq:CG_expansion_d}
\end{empheq}
\end{subequations}

\noindent Here, the symbol $\approx$ denotes truncation after the first grouped correction level. These expressions do not require the assumption $z_{1,0}\gg z_{2,0}$. Higher grouped corrections involve higher derivatives of the CG and higher powers and products of the oscillation amplitudes. Their complete expressions, together with the corresponding truncation regimes and numerical convergence analysis, are provided in the companion manuscript~\cite{ValloireSubmittedCG}.

To rationalize the dependence of the coefficients on the oscillation amplitudes, three truncation regimes of the CG series expansion are distinguished: the zeroth-order truncation regime (ZOTR), the first-order truncation regime (FOTR), and the higher-order truncation regime (HOTR). The full derivations, truncation regimes, and numerical convergence analysis of $K_i$ are provided in the companion manuscript~\cite{ValloireSubmittedCG}.

\clearpage

\section{Components of the electrostatic force influencing the bimodal dynamics}
\label{sec:electrostatic_force_components_bimodal_dynamics}

As mentioned in the introduction, AM-He-KPFM relies on a frequency-conversion scheme~\cite{Sugawara2012a}. The modulation angular frequency of the AC bias voltage, $\omega_{\mathrm{mod}}$ in Eq.~\eqref{eq:electrostatic_force_b}, is set to match the difference between the eigenmode angular frequencies, $\omega_{\mathrm{mod}}=\omega_2-\omega_1$. This seemingly simple operation requires a dedicated lock-in amplifier (LIA) equipped with multiple oscillators and sideband-tracking capabilities. The mechanical drive signal at frequency $\omega_1$, generated by the AFM controller, is tracked by a phase-locked loop (PLL). The output of the oscillator locked at $\omega_1$ is then mixed with the output of a second internal oscillator, whose frequency is set by the user close to the second eigenmode resonance, $\omega_2$. This enables the generation of reference signals at the mixed frequencies $(\omega_2\pm\omega_1)$, which are used to generate the AC bias voltage modulation.

When the bias modulation is performed accordingly, the electrostatic force contains sideband components at the eigenmode frequencies. Their first effect is to act as electromechanical power sources for the cantilever. These components arise from the electromechanical coupling between the cantilever dynamics and the bias-modulated electrostatic field. Their expressions can be derived by inserting the CG expansion, Eq.~\eqref{eq:CG_expansion}, into the modulated electrostatic force components, Eq.~\eqref{eq:electrostatic_components}.

The explicit calculation of all electrostatic force components is detailed in Sec.~SI-II. In what follows, the main results are summarized.

Eq.~\eqref{eq:electrostatic_components_b}, which yields $\omega_{\mathrm{mod}}$-induced sidebands, is first considered, see Sec.~SI-II~C. The calculation of the terms at $\omega_1$ and $\omega_2$ gives:

\begin{subequations}\label{eq:electrostatic_intermodulation_forces}
\begin{empheq}[left=\empheqlbrace]{align}
F_{\mathrm{el,int.}}^{\mathrm{dir.}}(t,z_c)
&= \alpha_1(z_c)\,z_{1,0}
\cos\!\left(\omega_2 t+\Phi_{\mathrm{mod}}+\Phi_1\right),
\label{eq:electrostatic_intermodulation_forces_a}
\\
F_{\mathrm{el,int.}}^{\mathrm{inv.}}(t,z_c)
&= \alpha_2(z_c)\,z_{2,0}
\cos\!\left(\omega_1 t-\Phi_{\mathrm{mod}}+\Phi_2\right),
\label{eq:electrostatic_intermodulation_forces_b}
\end{empheq}
\end{subequations}

\noindent where:

\begin{equation}\label{eq:alpha_coefficients}
\alpha_{1,2}(z_c)
=
\left|
\frac{K_{1,2}(z_c)}{2}
\left(V_{\mathrm{DC}}-V_{\mathrm{cpd}}\right)
U_{\mathrm{mod}}
\right|.
\end{equation}

In Eqs.~\eqref{eq:electrostatic_intermodulation_forces_a} and~\eqref{eq:electrostatic_intermodulation_forces_b}, the superscripts indicate that the oscillation of the first eigenmode acts, through heterodyne coupling and the resulting intermodulation products, as a power source for the second eigenmode, corresponding to the direct effect, ``dir.'', in Eq.~\eqref{eq:electrostatic_intermodulation_forces_a}, or conversely that the second eigenmode acts as a power source for the first one, corresponding to the inverse effect, ``inv.'', in Eq.~\eqref{eq:electrostatic_intermodulation_forces_b}. The subscript ``int.'' recalls that these terms originate from heterodyne intermodulation. These results explicitly show that nonlinear energy transfer between eigenmodes can occur. Such an effect was reported earlier~\cite{Kiracofe2013a,An2014a}, although without analytical support and outside the context of electrostatic interactions.

All other $\omega_{\mathrm{mod}}$-induced sidebands, namely electrostatic force components at frequencies different from $\omega_1$ or $\omega_2$, do not contribute to the conservative or non-conservative processes influencing the first two cantilever eigenmodes.

The coefficients $\alpha_i(z_c)$, where $i=1,2$, appearing in Eqs.~\eqref{eq:electrostatic_intermodulation_forces_a} and~\eqref{eq:electrostatic_intermodulation_forces_b}, represent $z_c$-dependent amplitudes of the electrostatic force and have the physical dimension of stiffness, [N/m]. For this reason, they are defined as positive quantities through the absolute value in Eq.~\eqref{eq:alpha_coefficients}. Since the coefficients $K_{1,2}(z_c)$ are themselves always positive, as shown in the companion manuscript~\cite{ValloireSubmittedCG}, this implies that, depending on the sign of $V_{\mathrm{DC}}-V_{\mathrm{cpd}}$, an additional $\pi$ term may enter the phase of the electrostatic force. In other words, $\Phi_{\mathrm{mod}}$ is effectively defined modulo $\pi$. It is also important to note that these coefficients vanish when either $V_{\mathrm{DC}}=V_{\mathrm{cpd}}$ or $U_{\mathrm{mod}}=0$.

The $2\omega_{\mathrm{mod}}$-induced sidebands derived from Eq.~\eqref{eq:electrostatic_components_c} can be safely neglected in the present analysis, see Sec.~SI-II~D.

Conversely, Eq.~\eqref{eq:electrostatic_components_a} yields additional electrostatic components at $\omega_1$ and $\omega_2$, arising from the bimodal cantilever dynamics within the static, or ``DC'', part of the electric field, see Sec.~SI-II~B. These components are referred to as DC-modulated electrostatic force components and are written as:

\begin{subequations}\label{eq:dc_modulated_electrostatic_forces}
\begin{empheq}[left=\empheqlbrace]{align}
F_{\mathrm{el,mod.}}^{\mathrm{DC},1}(t,z_c)
&= \beta_1(z_c)\,z_{1,0}
\cos\!\left(\omega_1 t+\Phi_1\right),
\label{eq:dc_modulated_electrostatic_forces_a}
\\
F_{\mathrm{el,mod.}}^{\mathrm{DC},2}(t,z_c)
&= \beta_2(z_c)\,z_{2,0}
\cos\!\left(\omega_2 t+\Phi_2\right),
\label{eq:dc_modulated_electrostatic_forces_b}
\end{empheq}
\end{subequations}

\noindent where:

\begin{equation}\label{eq:beta_coefficients}
\beta_{1,2}(z_c)
=
\frac{K_{1,2}(z_c)}{2}
\left[
\left(V_{\mathrm{DC}}-V_{\mathrm{cpd}}\right)^2
+\frac{U_{\mathrm{mod}}^2}{2}
\right].
\end{equation}

Because these components stem from the direct modulation of the electrostatic force, and not from an intermodulation product, they are denoted with the simple subscript ``mod.''. The coefficients $\beta_i(z_c)$, where $i=1,2$, are always positive and therefore do not affect the phase of these force components. In contrast to the coefficients $\alpha_i(z_c)$, however, they vanish only when both $V_{\mathrm{DC}}=V_{\mathrm{cpd}}$ and $U_{\mathrm{mod}}=0$. Finally, the coefficients $\beta_i(z_c)$ have the same physical dimension as $\alpha_i(z_c)$, namely stiffness, [N/m].

\clearpage

\section{Inter-mode energy exchange and coupled observables}
\label{sec:inter_mode_energy_exchange_coupled_observables}

\subsection{Parameters and observables}
\label{sec:parameters_and_observables}

The total force $F_{\mathrm{tot}}(t,z_c)$, see Eq.~\eqref{eq:sho_motion_c}, relevant to the projections at $\omega_1$ and $\omega_2$ and governing the bimodal dynamics of the cantilever is now fully established:

\begin{align}
F_{\mathrm{tot}}(t,z_c)
&= F_d(t)+F_{\mathrm{int}}(t,z_c)+F_{\mathrm{el}}(t,z_c)
\notag\\
&= F_d(t)+F_{\mathrm{int}}(t,z_c)
+F_{\mathrm{el,int.}}^{\mathrm{dir.}}(t,z_c)
+F_{\mathrm{el,int.}}^{\mathrm{inv.}}(t,z_c)
\notag\\
&\quad
+F_{\mathrm{el,mod.}}^{\mathrm{DC},1}(t,z_c)
+F_{\mathrm{el,mod.}}^{\mathrm{DC},2}(t,z_c).
\label{eq:total_force_components}
\end{align}

All the elements required to analyze the impact of the heterodyne process on the observables are now in place. For each eigenmode, this can be achieved by inserting the total force expression into the virial and power-balance equations, Eqs.~\eqref{eq:coupled_observables_a} and~\eqref{eq:coupled_observables_b}. The detailed calculations are provided in Sec.~SI-III, while the main results are presented here.

For nc-AFM AM-He-KPFM, the amplitude $z_{1,0}$ and phase $\Phi_1$ of the first eigenmode are kept constant and satisfy the resonance condition, such that $\Phi_1=-\pi/2$. The self-consistent description of the problem then involves four experimental observables and a set of user-defined parameters. The observables are:

\begin{itemize}
    \item for the first eigenmode:
    \begin{itemize}
            \item its frequency shift, $\Delta\omega_1$;
            \item the driving force, $F_d$, required to keep $z_{1,0}$ constant;
    \end{itemize}

    \item for the second eigenmode:
    \begin{itemize}
            \item its oscillation amplitude, $z_{2,0}$;
            \item its phase, $\Phi_2$, which is not locked.
    \end{itemize}
\end{itemize}

\noindent The user-defined parameters are:

\begin{itemize}
    \item the average tip-surface distance, $z_c$, controlled by the setpoint $\Delta\omega_1$ when the $z$-feedback loop is closed;
    \item the oscillation amplitude of the first eigenmode, $z_{1,0}$;
    \item the electrostatic modulation conditions, $V_{\mathrm{DC}}$, $U_{\mathrm{mod}}$, and $\Phi_{\mathrm{mod}}$;
    \item the angular frequency $\omega_2$, which both sets the demodulation frequency for $z_{2,0}$ and $\Phi_2$ and defines the modulation excitation frequency of $V_{\mathrm{mod}}(t)$ through the relation $\omega_{\mathrm{mod}}=\omega_2-\omega_1$.
\end{itemize}

\subsection{Expressions of the observables}
\label{sec:expressions_of_the_observables}

The virial and power-balance equations for the second eigenmode are first addressed. The calculations yield, see Sec.~SI-III~A:

\begin{subequations}\label{eq:second_mode_observables}
\begin{empheq}[left=\empheqlbrace]{align}
\frac{\Delta\omega_2(z_c)}{\omega_{2,0}}
&=
-\frac{k_{\mathrm{int},2}^{(c)}(z_c)}{k_2}
-\frac{\beta_2(z_c)}{2k_2}
-\frac{\alpha_1(z_c)}{2k_2}
\frac{z_{1,0}}{z_{2,0}}
\cos\!\left(\Phi_{\mathrm{mod}}+\Phi_1-\Phi_2\right),
\label{eq:second_mode_observables_a}
\\
\frac{\omega_2(z_c)}{2Q_2\omega_{2,0}}
&=
-\frac{k_{\mathrm{int},2}^{(d)}(z_c)}{k_2}
+\frac{\alpha_1(z_c)}{2k_2}
\frac{z_{1,0}}{z_{2,0}}
\sin\!\left(\Phi_{\mathrm{mod}}+\Phi_1-\Phi_2\right),
\label{eq:second_mode_observables_b}
\end{empheq}
\end{subequations}

\noindent where the non-electrostatic conservative, superscript $(c)$, and dissipative, superscript $(d)$, interaction-induced stiffness-like quantities acting on the second eigenmode are defined as:

\begin{subequations}\label{eq:second_mode_interaction_stiffnesses}
\begin{empheq}[left=\empheqlbrace]{align}
k_{\mathrm{int},2}^{(c)}(z_c)
&=
\frac{1}{z_{2,0}T_s}
\int_0^{T_s}
F_{\mathrm{int}}(t,z_c)
\cos\!\left(\omega_2 t+\Phi_2\right)\,dt,
\label{eq:second_mode_interaction_stiffnesses_a}
\\
k_{\mathrm{int},2}^{(d)}(z_c)
&=
\frac{1}{z_{2,0}T_s}
\int_0^{T_s}
F_{\mathrm{int}}(t,z_c)
\sin\!\left(\omega_2 t+\Phi_2\right)\,dt.
\label{eq:second_mode_interaction_stiffnesses_b}
\end{empheq}
\end{subequations}

An important phase condition can be deduced from the virial equation, Eq.~\eqref{eq:second_mode_observables_a}. In AM-He-KPFM, the second-eigenmode resonance frequency, $\omega_{2,0}$, is imposed by the cantilever and is only slightly shifted upon interaction with the surface. From an instrumental standpoint, however, the frequency at which the LIA demodulates the signal, denoted by $\omega_2$ in the present problem, is usually set to match $\omega_{2,0}$. In practice, $\omega_2$ can be tuned at will, whether or not it matches the actual resonance frequency of the second eigenmode. Tuning $\omega_2$ implicitly tunes the phase $\Phi_2$ of the eigenmode. This can be achieved by imposing the following quadrature phase relationship:

\begin{equation}\label{eq:quadrature_phase_condition}
\Phi_{\mathrm{mod}}+\Phi_1-\Phi_2
=
\pm\frac{\pi}{2}.
\end{equation}

Then, Eq.~\eqref{eq:second_mode_observables_a} becomes:

\begin{equation}\label{eq:second_mode_frequency_shift_quadrature}
\frac{\Delta\omega_2(z_c)}{\omega_{2,0}}
=
-\frac{1}{k_2}
\left[
k_{\mathrm{int},2}^{(c)}(z_c)
+\frac{\beta_2(z_c)}{2}
\right].
\end{equation}

Eq.~\eqref{eq:second_mode_frequency_shift_quadrature} describes the interaction-induced shift of the second-eigenmode resonance frequency upon approach to the surface. It shows that the conservative contribution of the heterodyne intermodulation term vanishes when the electrostatic excitation is applied exactly at the resonance of the second eigenmode. In practice, however, as mentioned above, $\omega_2$ may differ from the exact interaction-shifted resonance frequency. In the limit of small deviations, Eq.~\eqref{eq:quadrature_phase_condition} can be recast into a ``quasi-quadrature'' phase condition:

\begin{equation}\label{eq:quasi_quadrature_phase_condition}
\Phi_{\mathrm{mod}}+\Phi_1-\Phi_2
=
+\left(\frac{\pi}{2}-\varepsilon\right),
\qquad
|\varepsilon|\ll 1.
\end{equation}

The choice of a positive sign for the quasi-quadrature condition is dictated by physical consistency, as discussed below. By also applying the quasi-quadrature condition to the power-balance equation, the set of coupled equations governing the dynamics of the second eigenmode is obtained:

\begin{subequations}\label{eq:second_mode_dynamics_quasi_quadrature}
\begin{empheq}[left=\empheqlbrace]{align}
\frac{\Delta\omega_2(z_c)}{\omega_{2,0}}
&=
-\frac{k_{\mathrm{int},2}^{(c)}(z_c)}{k_2}
-\frac{\beta_2(z_c)}{2k_2}
-\frac{\alpha_1(z_c)}{2k_2}
\frac{z_{1,0}}{z_{2,0}}
\sin(\varepsilon),
\label{eq:second_mode_dynamics_quasi_quadrature_a}
\\
\frac{\omega_2(z_c)}{2Q_2\omega_{2,0}}
&=
-\frac{k_{\mathrm{int},2}^{(d)}(z_c)}{k_2}
+\frac{\alpha_1(z_c)}{2k_2}
\frac{z_{1,0}}{z_{2,0}}
\cos(\varepsilon).
\label{eq:second_mode_dynamics_quasi_quadrature_b}
\end{empheq}
\end{subequations}

In Eq.~\eqref{eq:second_mode_dynamics_quasi_quadrature_b}, the first term on the right-hand side accounts for other possible tip-surface dissipative processes. Owing to sign consistency, with a positive left-hand side, this term must be balanced by a second term that remains positive. This second term contains the oscillation amplitudes of the first and second eigenmodes, $z_{1,0}$ and $z_{2,0}>0$, as well as the coefficient $\alpha_1(z_c)>0$, see Eq.~\eqref{eq:alpha_coefficients}, which is proportional to $U_{\mathrm{mod}}$. To keep this contribution positive, $\cos(\varepsilon)$ must also be positive, which justifies the positive sign chosen for the quasi-quadrature phase condition in Eq.~\eqref{eq:quasi_quadrature_phase_condition}. Thus, the oscillation of the second eigenmode is driven by that of the first eigenmode within the modulated electric field. This well-known mechanism forms the basis of AM-He-KPFM and is referred to as the ``direct heterodyne'' effect.

The virial and power-balance equations for the first eigenmode can also be solved, see Sec.~SI-III~B. Upon applying the quasi-quadrature condition, Eq.~\eqref{eq:quasi_quadrature_phase_condition}, the following expressions are obtained:

\begin{subequations}\label{eq:first_mode_dynamics_quasi_quadrature}
\begin{empheq}[left=\empheqlbrace]{align}
\frac{\Delta\omega_1(z_c)}{\omega_{1,0}}
&=
-\frac{k_{\mathrm{int},1}^{(c)}(z_c)}{k_1}
-\frac{\beta_1(z_c)}{2k_1}
-\frac{\alpha_2(z_c)}{2k_1}
\frac{z_{2,0}}{z_{1,0}}
\sin(\varepsilon),
\label{eq:first_mode_dynamics_quasi_quadrature_a}
\\
\frac{\omega_1(z_c)}{2Q_1\omega_{1,0}}
&=
\frac{F_d}{2z_{1,0}k_1}
-\frac{k_{\mathrm{int},1}^{(d)}(z_c)}{k_1}
-\frac{\alpha_2(z_c)}{2k_1}
\frac{z_{2,0}}{z_{1,0}}
\cos(\varepsilon),
\label{eq:first_mode_dynamics_quasi_quadrature_b}
\end{empheq}
\end{subequations}

\noindent where the non-electrostatic conservative, superscript $(c)$, and dissipative, superscript $(d)$, interaction-induced stiffness-like quantities acting on the first eigenmode have been introduced:

\begin{subequations}\label{eq:first_mode_interaction_stiffnesses}
\begin{empheq}[left=\empheqlbrace]{align}
k_{\mathrm{int},1}^{(c)}(z_c)
&=
\frac{1}{z_{1,0}T_s}
\int_0^{T_s}
F_{\mathrm{int}}(t,z_c)
\cos\!\left(\omega_1 t+\Phi_1\right)\,dt,
\label{eq:first_mode_interaction_stiffnesses_a}
\\
k_{\mathrm{int},1}^{(d)}(z_c)
&=
\frac{1}{z_{1,0}T_s}
\int_0^{T_s}
F_{\mathrm{int}}(t,z_c)
\sin\!\left(\omega_1 t+\Phi_1\right)\,dt.
\label{eq:first_mode_interaction_stiffnesses_b}
\end{empheq}
\end{subequations}

As discussed at the beginning of this section, the four observables $\Delta\omega_1$, $F_d$, $z_{2,0}$, and $\Phi_2$, which are representative of nc-AFM open-loop AM-He-KPFM, are sought. The coupled nonlinear system to be considered is given by Eqs.~\eqref{eq:second_mode_dynamics_quasi_quadrature_a},~\eqref{eq:second_mode_dynamics_quasi_quadrature_b},~\eqref{eq:first_mode_dynamics_quasi_quadrature_a}, and~\eqref{eq:first_mode_dynamics_quasi_quadrature_b}. An exact solution of this system is beyond the scope of the present work. In the SI, an approximate solution based on linear approximations and on the self-consistency of the problem is proposed, see Sec.~SI-III~C. This corresponds to the viewpoint that all variables of the problem are determined once the average tip-surface distance $z_c$ is set. Since $z_c$ is a key user-defined parameter, the observables are expressed in a form that makes their dependence on $z_c$ as explicit as possible, while removing it from the remaining parameters. The final expressions of the observables as functions of the parameters are given below, see Sec.~SI-III~C~3:

\begin{subequations}\label{eq:open_loop_observables}
\begin{empheq}[left=\empheqlbrace]{align}
\frac{\Delta\omega_1(z_c)}{\omega_{1,0}}
&=
-\frac{1}{k_1}
\biggl[
k_{\mathrm{int},1}^{(c)}(z_c)
+\frac{\beta_1(z_c)}{2}
\notag\\
&\qquad
+\frac{\alpha_1(z_c)\alpha_2(z_c)}{2}
\widetilde{k}_2(z_c)
\left|\widehat{\widetilde{G}}_2(\omega_2,z_c)\right|^2
\left(1-u_2^2(\omega_2,z_c)\right)
\biggr],
\label{eq:open_loop_observables_a}
\\[0.5em]
F_d(z_c)
&=
z_{1,0}
\biggl[
\frac{\widetilde{k}_1(z_c)}{Q_1}
+2k_{\mathrm{int},1}^{(d)}(z_c)
\notag\\
&\qquad
+\alpha_1(z_c)\alpha_2(z_c)
\frac{\widetilde{k}_2(z_c)}{Q_2}
\left|\widehat{\widetilde{G}}_2(\omega_2,z_c)\right|^2
\biggr],
\label{eq:open_loop_observables_b}
\\[0.5em]
z_{2,0}(z_c)
&=
z_{1,0}\alpha_1(z_c)
\left|\widehat{\widetilde{G}}_2(\omega_2,z_c)\right|,
\label{eq:open_loop_observables_c}
\\
\Phi_2(z_c)
&=
\Phi_{\mathrm{mod}}+\Phi_1
+\Phi_{\widehat{\widetilde{G}}_2}(\omega_2,z_c).
\label{eq:open_loop_observables_d}
\end{empheq}
\end{subequations}

In the above equations, all quantities have already been defined, except $\widetilde{k}_i(z_c)$, $\left|\widehat{\widetilde{G}}_2(\omega_2,z_c)\right|$, and $\Phi_{\widehat{\widetilde{G}}_2}(\omega_2,z_c)$. The quantities $\widetilde{k}_i(z_c)$, with $i=1,2$, are effective stiffness-like quantities associated with the first and second eigenmodes. They are defined as follows, see Sec.~SI-III~C:

\begin{equation}\label{eq:effective_stiffness}
\widetilde{k}_i(z_c)
=
k_i
-
k_{\mathrm{int},i}^{(c)}(z_c)
-
\frac{\beta_i(z_c)}{2}.
\end{equation}

The quantity $u_2(\omega_2,z_c)$ is defined as follows, see Sec.~SI-III~C:

\begin{equation}\label{eq:normalized_second_mode_frequency}
u_2(\omega_2,z_c)
=
\frac{k_2}{\widetilde{k}_2(z_c)}
\frac{\omega_2}{\omega_{2,0}}
=
\frac{\omega_2}{\widetilde{\omega}_{2,0}(z_c)},
\end{equation}

\noindent where $\widetilde{\omega}_{2,0}(z_c)$ denotes the linearized interaction-shifted resonance frequency of the second eigenmode, which is defined as, see Sec.~SI-III~C:

\begin{equation}\label{eq:linearized_shifted_second_mode_resonance}
\widetilde{\omega}_{2,0}(z_c)
=
\frac{\widetilde{k}_2(z_c)}{k_2}
\omega_{2,0}.
\end{equation}

The quantities $\widetilde{k}_i(z_c)$, with $i=1,2$, are defined within the linearized frequency-shift formalism and include conservative and bias-modulation contributions at the average tip-surface distance $z_c$. However, they should not be interpreted as exact mechanical stiffnesses entering the oscillator relation $\omega=\sqrt{k/m}$. For an exact effective stiffness $k_i^{\mathrm{eff}}$, the resonance frequency would instead scale as $\widetilde{\omega}_{i,0}=\omega_{i,0}\sqrt{k_i^{\mathrm{eff}}/k_i}$.

The quantities $\left|\widehat{\widetilde{G}}_2(\omega_2,z_c)\right|$ and $\Phi_{\widehat{\widetilde{G}}_2}(\omega_2,z_c)$ denote the modulus and the phase of the transfer function of the interaction-shifted second eigenmode. Its resonance frequency and effective stiffness are $\widetilde{\omega}_{2,0}(z_c)$ and $\widetilde{k}_2(z_c)$, respectively, see Sec.~SI-III~C. These quantities are evaluated at the angular frequency $\omega_2$, see Eqs.~\eqref{eq:transfer_function_b} and~\eqref{eq:transfer_function_c}:

\begin{subequations}\label{eq:shifted_second_mode_transfer_function}
\begin{empheq}[left=\empheqlbrace]{align}
\left|\widehat{\widetilde{G}}_2(\omega_2,z_c)\right|
&=
\frac{1}{\widetilde{k}_2(z_c)}
\frac{1}{
\sqrt{
\left(
\frac{\omega_2}{Q_2\widetilde{\omega}_{2,0}(z_c)}
\right)^2
+
\left[
1-
\left(
\frac{\omega_2}{\widetilde{\omega}_{2,0}(z_c)}
\right)^2
\right]^2
}
},
\label{eq:shifted_second_mode_transfer_function_a}
\\
\Phi_{\widehat{\widetilde{G}}_2}(\omega_2,z_c)
&=
-\operatorname{atan2}\!\left(
\frac{\omega_2}{\widetilde{\omega}_{2,0}(z_c)Q_2},
1-\frac{\omega_2^2}{\widetilde{\omega}_{2,0}^2(z_c)}
\right).
\label{eq:shifted_second_mode_transfer_function_b}
\end{empheq}
\end{subequations}

\subsection{Consequences}
\label{sec:consequences}

The consequences of the above equations for the observables measured during nc-AFM open-loop AM-He-KPFM experiments are now discussed. The predictions are compared with experimental results in the next section.

The results of Sec.~\ref{sec:expressions_of_the_observables} are first summarized. Eq.~\eqref{eq:effective_stiffness} describes how the effective stiffness of each eigenmode is modified within the present framework. It shows that the intrinsic stiffness $k_i$ of each eigenmode, with $i=1,2$, is reduced by the conservative interaction contribution, $k_{\mathrm{int},i}^{(c)}(z_c)$, and by the DC-modulated electrostatic contribution, $\beta_i(z_c)/2$.

Eq.~\eqref{eq:linearized_shifted_second_mode_resonance} provides a linear description of the interaction-shifted resonance frequency of the second eigenmode, $\omega_{2,0}\rightarrow\widetilde{\omega}_{2,0}(z_c)$, induced by the corresponding stiffness change, $k_2\rightarrow\widetilde{k}_2(z_c)$.

The quantity $\left|\widehat{\widetilde{G}}_2(\omega_2,z_c)\right|$, see Eq.~\eqref{eq:shifted_second_mode_transfer_function_a}, is the modulus of the transfer function of the interaction-shifted second eigenmode. It reaches its maximum when $\omega_2=\widetilde{\omega}_{2,0}(z_c)$ and decreases around this value over a characteristic width of approximately $\widetilde{\omega}_{2,0}(z_c)/Q_2$. The quantity $\Phi_{\widehat{\widetilde{G}}_2}(\omega_2,z_c)$, see Eq.~\eqref{eq:shifted_second_mode_transfer_function_b}, is the corresponding phase of the interaction-shifted second eigenmode, evaluated on the continuous oscillator branch using the two-argument arctangent. Substituting these expressions into the observable equations is useful to describe several experimental trends more accurately, as discussed in the next section.

These quantities enter the set of observables given by Eq.~\eqref{eq:open_loop_observables}, which helps clarify their physical meaning. Eq.~\eqref{eq:open_loop_observables_a} gives the frequency shift of the first eigenmode, as measured in nc-AFM experiments. As expected, this shift is negative and governed by the conservative part of the interaction, $k_{\mathrm{int},1}^{(c)}(z_c)$, but also by the DC-modulated electrostatic contribution contained in the term $\beta_1(z_c)/2\propto (V_{\mathrm{DC}}-V_{\mathrm{cpd}})^2+U_{\mathrm{mod}}^2/2$, see Eq.~\eqref{eq:beta_coefficients}. This term accounts for the well-known quadratic dependence of the frequency shift on $V_{\mathrm{DC}}$, and also on $U_{\mathrm{mod}}$~\cite{Sadewasser2012a,Sadewasser2018a}. The last term in the brackets is less expected. It is a signature of the inverse heterodyne effect, meaning that the heterodyne electrostatic modulation acting on the second eigenmode feeds back onto the first one. This term is proportional to $\alpha_1(z_c)\alpha_2(z_c)\propto (V_{\mathrm{DC}}-V_{\mathrm{cpd}})^2U_{\mathrm{mod}}^2$, see Eq.~\eqref{eq:alpha_coefficients}, showing that the quadratic dependence of the frequency shift on $V_{\mathrm{DC}}$ and $U_{\mathrm{mod}}$ is also affected by this mechanism. However, the strength of the inverse heterodyne contribution to the first-eigenmode frequency shift is expected to remain small, since it is weighted by the factor $\left|\widehat{\widetilde{G}}_2(\omega_2,z_c)\right|^2\left[1-u_2^2(\omega_2,z_c)\right]$, which is controlled by the demodulation frequency $\omega_2$. In most cases, demodulation is performed such that $\omega_2\approx\widetilde{\omega}_{2,0}(z_c)$, implying $u_2(\omega_2\approx\widetilde{\omega}_{2,0},z_c)\approx1$ and therefore $\left|\widehat{\widetilde{G}}_2(\omega_2\approx\widetilde{\omega}_{2,0},z_c)\right|^2\left[1-u_2^2(\omega_2\approx\widetilde{\omega}_{2,0},z_c)\right]\approx0$. This contribution also vanishes when $\omega_2$ is sufficiently far from $\widetilde{\omega}_{2,0}(z_c)$, because in that case $\left|\widehat{\widetilde{G}}_2(\omega_2,z_c)\right|\approx0$. Interestingly, in the vicinity of $\widetilde{\omega}_{2,0}(z_c)$, this term changes sign. When $\omega_2\lesssim\widetilde{\omega}_{2,0}(z_c)$, one has $u_2(\omega_2\lesssim\widetilde{\omega}_{2,0},z_c)\lesssim1$, see Eq.~\eqref{eq:normalized_second_mode_frequency}, and thus $\left|\widehat{\widetilde{G}}_2(\omega_2\lesssim\widetilde{\omega}_{2,0},z_c)\right|^2\left[1-u_2^2(\omega_2\lesssim\widetilde{\omega}_{2,0},z_c)\right]>0$. Conversely, $\left|\widehat{\widetilde{G}}_2(\omega_2\gtrsim\widetilde{\omega}_{2,0},z_c)\right|^2\left[1-u_2^2(\omega_2\gtrsim\widetilde{\omega}_{2,0},z_c)\right]<0$ when $\omega_2\gtrsim\widetilde{\omega}_{2,0}(z_c)$. Consequently, Eq.~\eqref{eq:open_loop_observables_a} predicts that, when $\omega_2\lesssim\widetilde{\omega}_{2,0}(z_c)$, the inverse heterodyne effect tends to increase the effective interaction stiffness contribution, making the first-eigenmode frequency shift algebraically smaller, i.e. more negative. Conversely, when $\omega_2\gtrsim\widetilde{\omega}_{2,0}(z_c)$, it tends to decrease this contribution, making the frequency shift algebraically larger, i.e. less negative. This behavior is therefore equivalent to an apparent decrease or increase, respectively, in the effective stiffness of the first eigenmode.

Eq.~\eqref{eq:open_loop_observables_b} gives the driving force that must be applied to the first eigenmode to keep its oscillation amplitude $z_{1,0}$ constant during nc-AFM experiments, and therefore describes its power balance, i.e. gain and loss. The first term in the brackets describes the power balance associated with the stiffness change $k_1\rightarrow\widetilde{k}_1(z_c)$. As discussed above, the conservative part of the interaction, together with the DC-modulated electrostatic contribution, tends to lower $F_d(z_c)$ through the terms $-k_{\mathrm{int},1}^{(c)}(z_c)-\beta_1(z_c)/2$ in the expression of $\widetilde{k}_1(z_c)$, see Eq.~\eqref{eq:effective_stiffness}. The second term in the brackets is expected and reflects the sensitivity of the driving force to tip-surface dissipative effects. The influence of conservative effects on $F_d(z_c)$ is expected to be negligible compared with that of dissipative effects because of the $1/Q_1$ weighting factor. The third term in the brackets, which is also a signature of the inverse heterodyne effect, is less expected. This term is proportional to $\alpha_1(z_c)\alpha_2(z_c)\propto (V_{\mathrm{DC}}-V_{\mathrm{cpd}})^2U_{\mathrm{mod}}^2$ and shows that an additional dissipation channel for the first eigenmode, corresponding to an increase of the driving force, arises from the electrostatic modulation. This effect is strongest when $\omega_2$ matches $\widetilde{\omega}_{2,0}(z_c)$, since $\left|\widehat{\widetilde{G}}_2(\omega_2,z_c)\right|^2$ is then maximal. A quadratic dependence of the observable $F_d(z_c)$ on $V_{\mathrm{DC}}$, and also on $U_{\mathrm{mod}}$, is therefore predicted as a consequence of the inverse heterodyne effect.

Eq.~\eqref{eq:open_loop_observables_c} gives the oscillation amplitude of the second eigenmode induced by the heterodyne electrostatic modulation. It is proportional to $z_{1,0}$, the oscillation amplitude of the first eigenmode, and to $\alpha_1(z_c)\propto\left|(V_{\mathrm{DC}}-V_{\mathrm{cpd}})U_{\mathrm{mod}}\right|$. One therefore expects a linear dependence of the observable on $z_{1,0}$, on $\left|V_{\mathrm{DC}}-V_{\mathrm{cpd}}\right|$, and on $U_{\mathrm{mod}}$. As a consequence of resonant detection, $z_{2,0}(z_c)$ is also expected to be maximum when $\left|\widehat{\widetilde{G}}_2(\omega_2,z_c)\right|$ is maximal, namely when $\omega_2=\widetilde{\omega}_{2,0}(z_c)$.

Finally, Eq.~\eqref{eq:open_loop_observables_d} gives the phase of the second eigenmode as a function of the demodulation angular frequency $\omega_2$. This expression results from the sum of the phase of the first-eigenmode deflection and that of the modulation voltage through frequency mixing, together with the phase associated with the transfer function of the second eigenmode. The latter equals $-\pi/2$ when $\omega_2=\widetilde{\omega}_{2,0}(z_c)$.

In the next section, the theoretical developments are compared with the experimental results.

\clearpage

\section{Experimental signatures of inverse heterodyne back-action}
\label{sec:experimental_signatures_inverse_heterodyne_back_action}

The experiments were performed in a UHV setup (base pressure $2\times10^{-10}$~mbar) with a beam-deflection VT-AFM microscope from Scienta Omicron~\cite{Scienta_Omicron} at room temperature. The instrument used here features an \textit{in situ} preamplifier that converts the photocurrent from the four-quadrant photodetector into a voltage signal, along with an \textit{ex situ} processing MathBox that computes the resulting vertical and lateral deflections of the cantilever. This setup provides a 3~MHz detection bandwidth for the cantilever deflection, which, with the probes used here, allows the first two eigenmodes of the cantilever to be detected without attenuation. The microscope is equipped with an RHK R9 electronics unit enabling nc-AFM operation~\cite{RHK_Technology}.

The cantilever is a PPP-NCL-Pt probe from NanoSensors with a backside Pt coating~\cite{Nanosensors}. Its parameters are reported in Table~\ref{tab:cantilever_parameters}. Except for the row ``stiffness'', which corresponds to the manufacturer's nominal specification, all reported values were obtained from experimental measurements. In particular, the oscillation amplitude of the first eigenmode, $z_{1,0}$, was carefully calibrated using the constant-$\gamma$ method, which is known to be reasonably accurate~\cite{giessibl00a,simon07a}, although recently reported methods appear to provide higher accuracy~\cite{heile21a}. For most of the experiments reported here, unless otherwise specified, $z_{1,0}=(12\pm0.5)$~nm was used. The oscillation amplitude of the second eigenmode, $z_{2,0}$, was not directly calibrated. Its calibration was instead implicitly deduced from the ratio between the eigenmode peaks at $f_{2,0}$ and $f_{1,0}$ in the Fourier spectrum of the cantilever deflection signal, as measured with the LIA used here. The second-eigenmode amplitude is estimated as $z_{2,0}=z_{1,0}/68=(0.18\pm0.02)$~nm. This calibration is not strictly accurate, since the two eigenmodes do not have the same detection sensitivity with the four-quadrant photodetector. Nevertheless, this calibration is considered to provide a reasonable order-of-magnitude estimate of the physical amplitude of the second eigenmode. In Table~\ref{tab:cantilever_parameters}, it is also noted that $f_{1,0}$ and $f_{2,0}\approx6.2f_{1,0}$ are nearly commensurate, yielding a super-frequency $f_s\approx25$~Hz.

The sample consists of an Al$_2$O$_3$-passivated p-doped crystalline silicon (c-Si) substrate, similar to the one described in Ref.~\cite{Aubriet2022a}. Owing to its optoelectronic properties, it can serve as a reference sample for photo-assisted KPFM experiments. In the present study, however, it is used only as a test substrate. Experiments under illumination will be reported elsewhere.

The experiments reported below were also reproduced on a metallic Au(111) sample, with qualitatively similar results. To keep both the main text and the SI as concise as possible, these data are not shown.

The nc-AFM mode provides access to the observables related to the first eigenmode: the drive force $F_d$, used to quantify non-conservative interactions through the power equation, see Eq.~\eqref{eq:open_loop_observables_b}, and the frequency shift $\Delta f_1=\Delta\omega_1/(2\pi)$, which reflects conservative interactions through the virial equation, see Eq.~\eqref{eq:open_loop_observables_a}. Experimentally, the quantity that is directly accessible is not $F_d$, but the drive voltage $V_d$, defined as the amplitude of the AC voltage applied to the piezotube used to excite the first eigenmode. Instead of the drive force $F_d$, $V_d$ is therefore used in the experimental figures and is assumed to be proportional to $F_d$. Likewise, the average tip-surface distance $z_c$ is not directly accessible experimentally. Instead, the measured quantity corresponds to the relative displacement of the piezotube to which the sample is attached, which controls the tip-surface separation. This quantity is referred to as the ``relative tip displacement'' in the experimental figures. As a result, $z_c$ is determined up to an additive offset set by the arbitrary zero of the piezotube displacement. Nevertheless, variations of $V_d$ and of the piezotube displacement remain physically meaningful, as they directly reflect variations of the drive force $F_d$ and of the average tip-surface distance $z_c$, respectively. For most experiments, unless otherwise stated, $\Delta f_1=-20$~Hz was used. These parameters enable a clearer observation of the effects discussed throughout this work.

The AM-He-KPFM configuration is implemented using an HF2LI LIA from Zurich Instruments, controlled with the LabOne software~\cite{Zurich_Instruments}. The electrostatic modulation parameters are listed in Table~\ref{tab:cantilever_parameters}. The HF2LI demodulates the cantilever deflection signal at $\omega_2$, near the resonance frequency of the second eigenmode, $\widetilde{\omega}_{2,0}(z_c)$, providing access to the third and fourth observables of the problem: $z_{2,0}$ and $\Phi_2$.

In the theoretical expressions, frequencies are written as angular frequencies. In the experimental discussion and in the figures, ordinary frequencies are used, with
$f_2=\omega_2/(2\pi)$, $f_{2,0}=\omega_{2,0}/(2\pi)$, and
$\widetilde{f}_{2,0}(z_c)=\widetilde{\omega}_{2,0}(z_c)/(2\pi)$.
The quantity $f_2$ denotes the demodulation frequency imposed by the LIA, while $f_{2,0}$ and $\widetilde{f}_{2,0}(z_c)$ denote the free and interaction-shifted second-eigenmode resonances, respectively.

In contrast to the preceding theoretical sections, the explicit dependence on $z_c$ is not systematically indicated in the experimental discussion. It is retained only when the dependence of the relevant quantities on $z_c$ is specifically discussed, in order to avoid ambiguity when presenting measurements as functions of other control parameters and to keep the notation lighter.

The observables depend on the following experimental parameters: $z_{1,0}$, $z_c$, $V_{\mathrm{DC}}$, $U_{\mathrm{mod}}$, and $f_2$. The amplitude setpoint $z_{1,0}$, selected by the user, is maintained at a constant value by a PID controller that adjusts $V_d$, and hence $F_d$, accordingly. The PLL maintains $\Delta f_1$ equal to a preset setpoint and, when the $z$-feedback loop is closed, this setpoint determines the average tip-surface distance $z_c$. The bias waveforms $V_{\mathrm{mod}}(t)$, including $U_{\mathrm{mod}}$, $\Phi_{\mathrm{mod}}$, and $f_{\mathrm{mod}}=f_2-f_1$, as well as $V_{\mathrm{DC}}$, are controlled and applied to the tip with the HF2LI. Before performing the measurements, the resonance of the second eigenmode is determined in order to adjust $f_2$ close to the resonance frequency $f_{2,0}$, see the parameters in Table~\ref{tab:cantilever_parameters}. The contact potential difference, $V_{\mathrm{cpd}}$, is then measured by recording $z_{2,0}(V_{\mathrm{DC}})$. Under the present experimental conditions, $V_{\mathrm{cpd}}=-200$~mV.

The four observables are acquired simultaneously while varying the five control parameters one by one. Except when varying $z_c$, the $z$-feedback loop is kept open during data acquisition to prevent any modification of the average tip-surface distance from affecting the signals. For each measurement, the acquisition time is kept sufficiently short to ensure that the microscope $z$-drift remains negligible and does not affect the data.

\subsection{Direct heterodyne effect: dependence of the second-eigenmode amplitude $z_{2,0}$ on $z_{1,0}$, $V_{\mathrm{DC}}$, $U_{\mathrm{mod}}$, and $z_c$}
\label{sec:direct_heterodyne_effect_z20}

\subsubsection{Evolution of $z_{2,0}$ with $z_{1,0}$}
\label{sec:z20_vs_z10}

\begin{figure}[htbp]
    \centering
    \includegraphics[width=\textwidth]{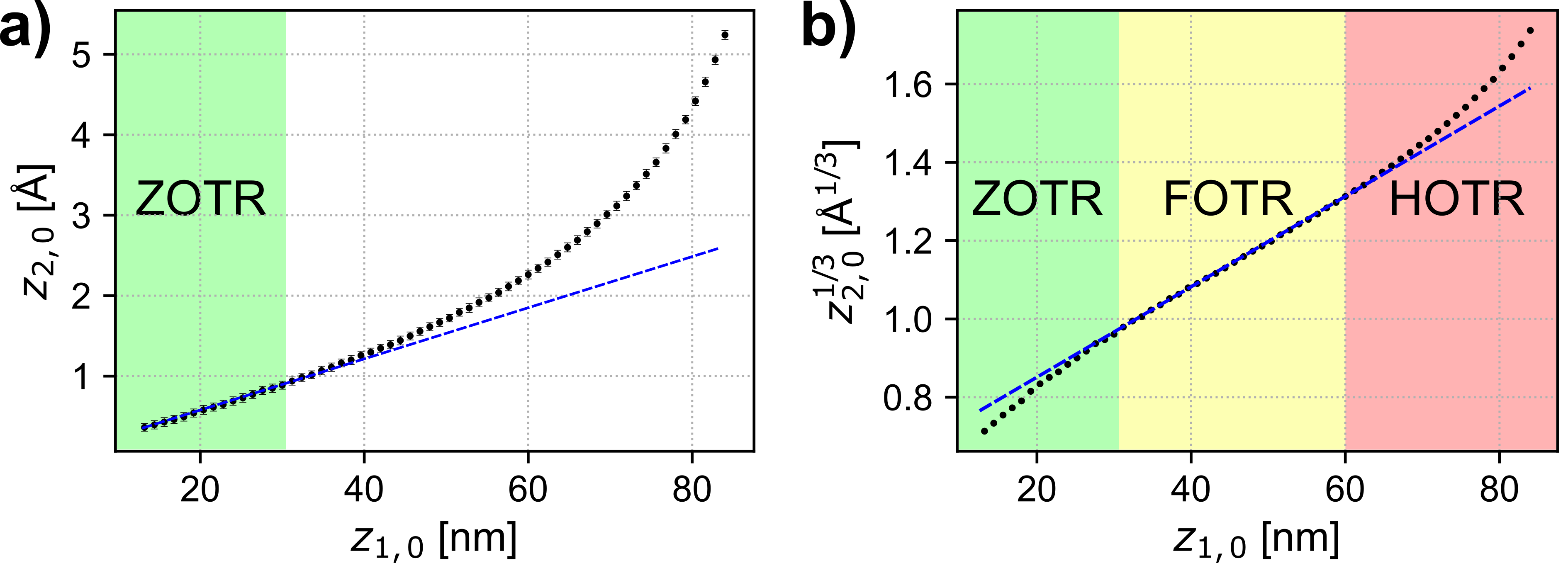}
    \caption{Experimental evolution of $z_{2,0}(z_{1,0})$, acquired with the $z$-feedback loop kept open. (a) $z_{2,0}(z_{1,0})$ increases linearly in the ZOTR ($z_{1,0}<30$~nm, green shaded area). The blue dashed line indicates the corresponding linear trend, from which clear deviations appear at larger $z_{1,0}$. (b) Same dataset replotted as $z_{2,0}^{1/3}(z_{1,0})$. A linear regime is observed in the FOTR for $z_{1,0}$ ranging from 30 to 60~nm (yellow shaded area and blue dashed line). For $z_{1,0}>60$~nm, the system enters the HOTR (red shaded area), where higher-order CG contributions become significant.}
    \label{fig:z20_vs_z10}
\end{figure}

Eq.~\eqref{eq:open_loop_observables_c} is first tested; it predicts that the oscillation amplitude of the second eigenmode, $z_{2,0}$, scales as $z_{1,0}\alpha_1$ and therefore as $z_{1,0}K_1$ under fixed electrostatic and demodulation conditions. To this end, $z_{2,0}(z_{1,0})$ was measured for $z_{1,0}$ values ranging from 12 to 84~nm, while keeping the $z$-feedback loop open so that the measurements were performed at constant $z_c$. Since $V_{\mathrm{DC}}$, $U_{\mathrm{mod}}$, and $f_2$ were also kept constant, the observed evolution directly probes the dependence of the effective CG coefficient $K_1$ on the oscillation amplitude. As shown in Fig.~\ref{fig:z20_vs_z10}(a), $z_{2,0}(z_{1,0})$ scales approximately linearly with $z_{1,0}$ for $z_{1,0}<30$~nm, which corresponds to the ZOTR, where $K_1\approx C^{(2)}(z_c)$ and higher-order terms in Eq.~\eqref{eq:CG_expansion_c} are negligible. For larger amplitudes, clear deviations from this linear trend appear, indicating that the FOTR correction $\left[z_{1,0}^{2}/8+z_{2,0}^{2}/4\right]C^{(4)}(z_c)$ becomes significant. Under the present experimental conditions, $z_{2,0}\ll z_{1,0}$, so that this correction is dominated by the contribution $C^{(4)}(z_c)z_{1,0}^{2}/8$. When this amplitude-dependent contribution becomes dominant over the leading term $C^{(2)}(z_c)$, Eq.~\eqref{eq:CG_expansion_c} yields $K_1\propto z_{1,0}^{2}$ and therefore $z_{2,0}(z_{1,0})\propto z_{1,0}^{3}$. This behavior is directly evidenced in Fig.~\ref{fig:z20_vs_z10}(b), where the same dataset plotted as $z_{2,0}^{1/3}(z_{1,0})$ exhibits a linear dependence between 30 and 60~nm, identifying a correction-dominated FOTR of $K_1$. At still larger amplitudes, for $z_{1,0}>60$~nm, the deviation from this cubic scaling reveals the onset of a HOTR of $K_1$, where higher-order derivatives of the CG become non-negligible. These measurements therefore provide a signature of the successive truncation regimes of $K_1$.

\subsubsection{Evolution of $z_{2,0}$ with $V_{\mathrm{DC}}$ and $U_{\mathrm{mod}}$}
\label{sec:z20_vs_vdc_umod}

\begin{figure}[htbp]
    \centering
    \includegraphics[width=\textwidth]{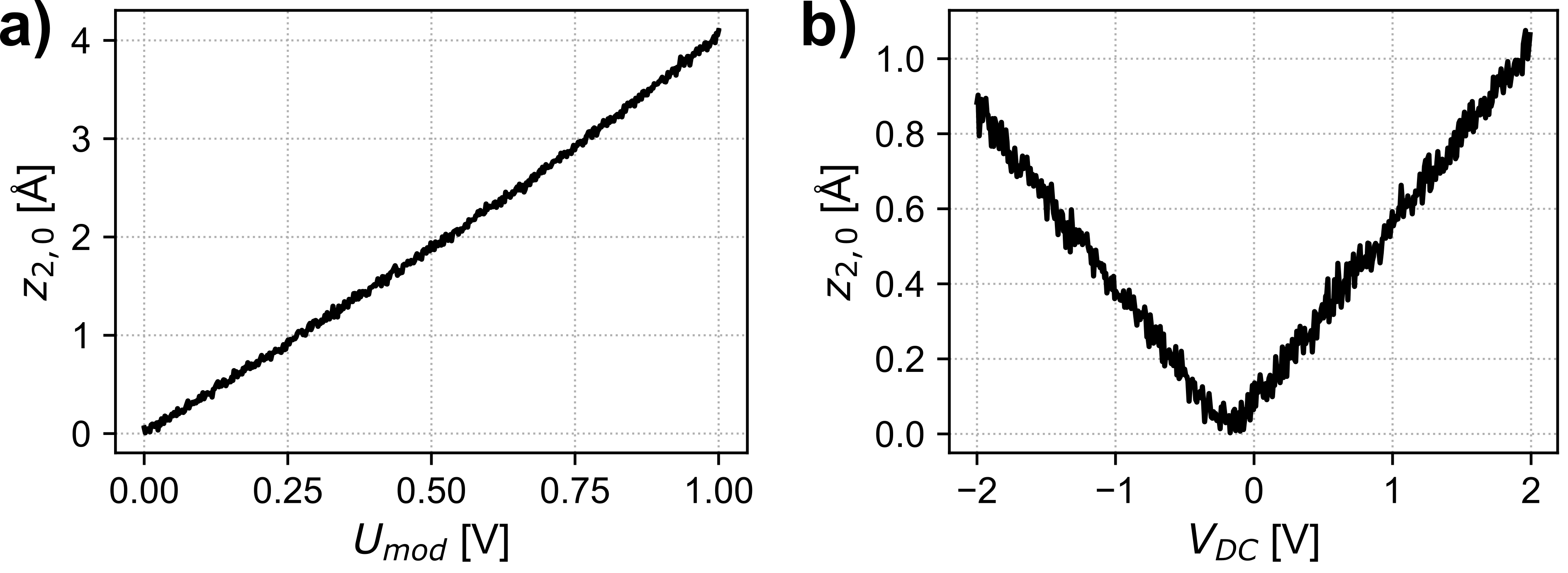}
    \caption{Evolution of $z_{2,0}(U_{\mathrm{mod}})$ (a) and $z_{2,0}(V_{\mathrm{DC}})$ (b), acquired with the $z$-feedback loop open. In agreement with Eq.~\eqref{eq:open_loop_observables_c}, $z_{2,0}$ scales linearly with $U_{\mathrm{mod}}$ and with $\left|V_{\mathrm{DC}}-V_{\mathrm{cpd}}\right|$. As expected, $z_{2,0}$ vanishes when $U_{\mathrm{mod}}=0$, or when $V_{\mathrm{DC}}=V_{\mathrm{cpd}}$.}
    \label{fig:z20_vs_umod_vdc}
\end{figure}

$z_{2,0}(U_{\mathrm{mod}})$ and $z_{2,0}(V_{\mathrm{DC}})$ were then measured. The results are reported in Fig.~\ref{fig:z20_vs_umod_vdc}(a) and Fig.~\ref{fig:z20_vs_umod_vdc}(b), respectively. The proportionality of $z_{2,0}$ to $U_{\mathrm{mod}}$ and $\left|\Delta V\right|=\left|V_{\mathrm{DC}}-V_{\mathrm{cpd}}\right|$ through the coefficient $\alpha_1$ is clearly observed, see Eqs.~\eqref{eq:alpha_coefficients} and~\eqref{eq:open_loop_observables_c}. When $V_{\mathrm{DC}}=V_{\mathrm{cpd}}$ or $U_{\mathrm{mod}}=0$, $z_{2,0}$ vanishes, as predicted, see Fig.~\ref{fig:z20_vs_umod_vdc}(b) or Fig.~\ref{fig:z20_vs_umod_vdc}(a). In contrast to the previous case, $z_{1,0}$ and $z_c$ are kept constant during the measurements, with amplitude regulation on $z_{1,0}$ and an open $z$-feedback loop. As a result, varying $U_{\mathrm{mod}}$ or $V_{\mathrm{DC}}$ alone does not significantly modify the CG regime, and no significant deviation from the linear dependence of $z_{2,0}$ on these parameters is observed.

\subsubsection{Evolution of $z_{2,0}$ with $z_c$}
\label{sec:z20_vs_zc}

\begin{figure}[htbp]
    \centering
    \includegraphics[width=\textwidth]{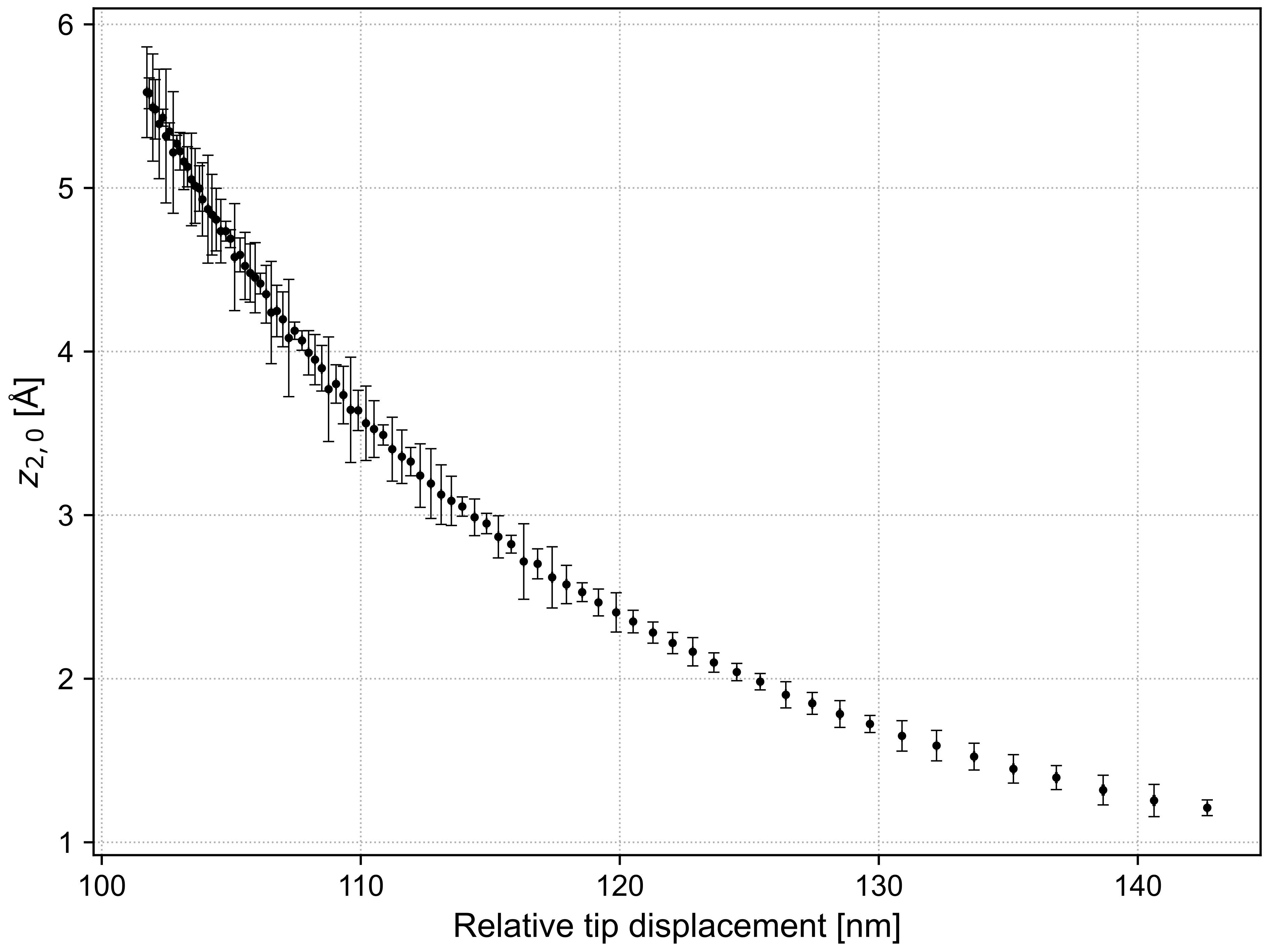}
    \caption{Evolution of $z_{2,0}$ as a function of the relative tip displacement ($z_c+\mathrm{offset}$), acquired with the $z$-feedback loop closed. Since $z_{2,0}(z_c)\propto\alpha_1(z_c)\propto K_1(z_c)$, see Eqs.~\eqref{eq:alpha_coefficients} and~\eqref{eq:open_loop_observables_c}, the evolution of $z_{2,0}(z_c)$ directly reflects the dependence of $K_1(z_c)$ on the average tip--surface distance. In the ZOTR, $K_1(z_c)\approx C^{(2)}(z_c)$, see Eq.~\eqref{eq:CG_expansion_c} and the companion manuscript, which is expected to decay with increasing $z_c$.}
    \label{fig:z20_vs_zc}
\end{figure}

To complete the analysis of the direct heterodyne effect and compare Eq.~\eqref{eq:open_loop_observables_c} with the experimental results, the dependence of $z_{2,0}$ on $z_c$ was also measured and is reported in Fig.~\ref{fig:z20_vs_zc}. These measurements were performed with the $z$-feedback loop closed, while $z_c$ was controlled through the $\Delta f_1$ setpoint. As already discussed, since $z_{2,0}(z_c)$ is proportional to $\alpha_1(z_c)\propto K_1(z_c)$, the behavior of $z_{2,0}(z_c)$ directly reflects that of $K_1(z_c)$. In the ZOTR, $K_1(z_c)\approx C^{(2)}(z_c)$ (see companion manuscript~\cite{ValloireSubmittedCG}), so this coefficient follows a decaying dependence on $z_c$. This trend is indeed observed for $z_{2,0}(z_c)$.

\subsection{Inverse heterodyne effect: dependence of the first-eigenmode observables $F_d$ and $\Delta f_1$ on $f_2$, $V_{\mathrm{DC}}$, and $U_{\mathrm{mod}}$}
\label{sec:inverse_heterodyne_effect_fd_df1}

The dependence of the experimental observables of the first eigenmode, $\Delta f_1$ and $F_d$, on $f_2$, $V_{\mathrm{DC}}$, and $U_{\mathrm{mod}}$ is now verified, see Eqs.~\eqref{eq:open_loop_observables_a} and~\eqref{eq:open_loop_observables_b}. It should be kept in mind that the $F_d$ signal, experimentally measured through $V_d$, and the $\Delta f_1$ signal are acquired simultaneously.

\subsubsection{Evolution of $F_d$ and $\Delta f_1$ with $f_2$}
\label{sec:fd_df1_vs_f2}

In Eqs.~\eqref{eq:open_loop_observables_a} and~\eqref{eq:open_loop_observables_b}, the influence of $f_2$ on $\Delta f_1$ and $F_d$ is directly related to the inverse heterodyne effect. This is expected, since the choice of $f_2$ affects both the amplitude and phase of the second-eigenmode oscillation through the cantilever transfer function $\widehat{\widetilde{G}}_2(f_2)$ at this mode.

\begin{figure}[htbp]
    \centering
    \includegraphics[width=\textwidth]{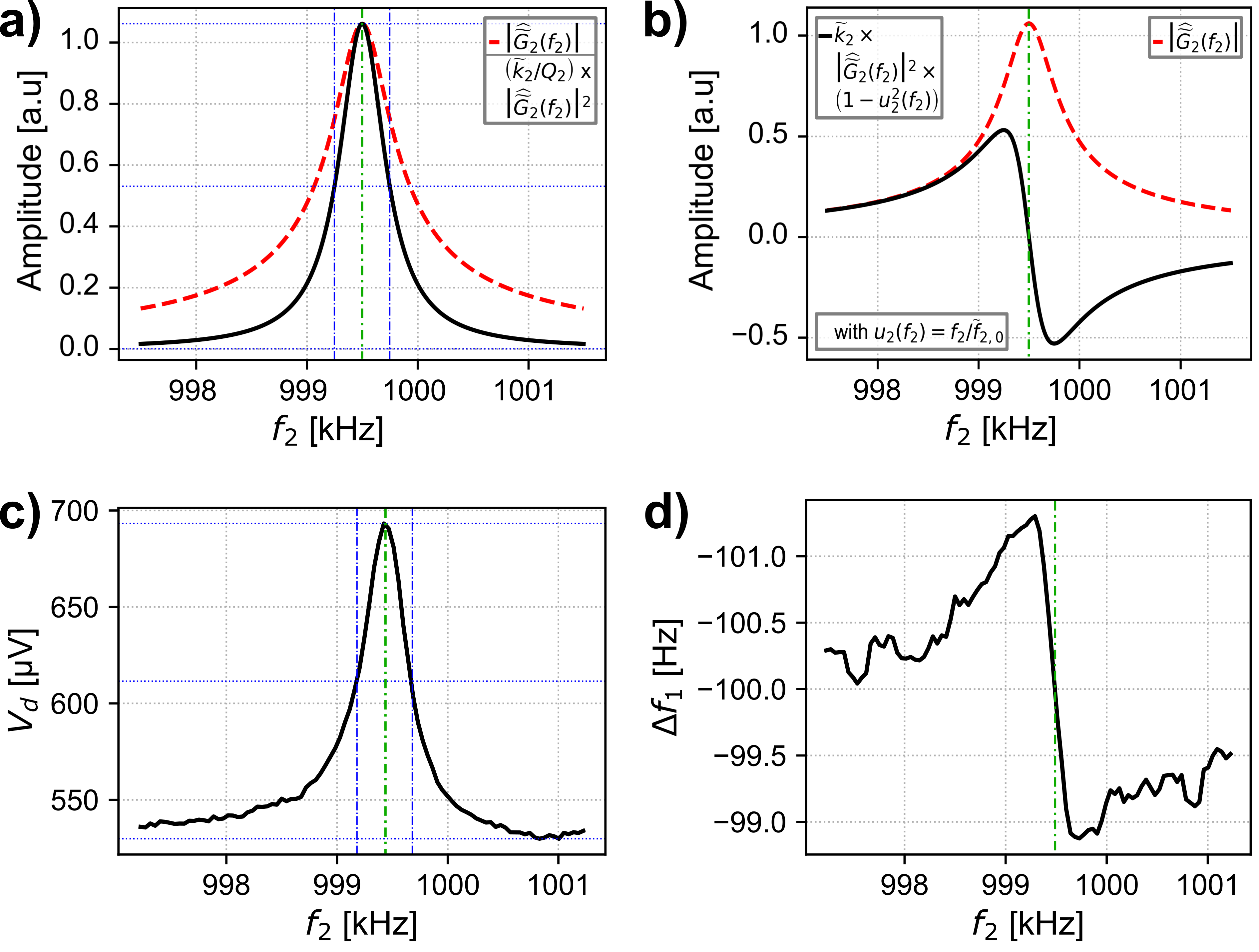}
    \caption{(a and b) Theoretical evolution of the terms $\left|\widehat{\widetilde{G}}_2(f_2)\right|^2$ (scaled by the prefactor $\widetilde{k}_2/Q_2$ for clarity) and $\left|\widehat{\widetilde{G}}_2(f_2)\right|^2\left[1-u_2^2(f_2)\right]$ (scaled by the prefactor $\widetilde{k}_2$ for clarity) as functions of $f_2$ (solid black curves), which appear in the expressions of $F_d$ and $\Delta f_1$, respectively, see Eqs.~\eqref{eq:open_loop_observables_b} and~\eqref{eq:open_loop_observables_a}. The numerical parameters were chosen to match the experimental ones, see Table~\ref{tab:cantilever_parameters}. The modulus of the transfer function, $\left|\widehat{\widetilde{G}}_2(f_2)\right|$, is also shown for comparison (dashed red curve). Squaring the transfer function leads, in the high-$Q$ limit, to a resonance peak whose full width at half maximum is $1/\sqrt{3}$ times that of the transfer-function modulus. The term $\left|\widehat{\widetilde{G}}_2(f_2)\right|^2\left[1-u_2^2(f_2)\right]$ changes sign at resonance and exhibits a positive lobe below and a negative lobe above the resonance frequency. (c and d) Experimental measurements of $V_d(f_2)$ and $\Delta f_1(f_2)$, acquired with the $z$-feedback loop open. The experimental curves closely follow the theoretical predictions, providing direct experimental evidence of the inverse heterodyne effect acting on the first eigenmode.}
    \label{fig:fd_df1_vs_f2}
\end{figure}

Fig.~\ref{fig:fd_df1_vs_f2}(a) and Fig.~\ref{fig:fd_df1_vs_f2}(b) show the theoretical $f_2$ dependence of the two transfer-function-related terms appearing in the expressions of $F_d$ and $\Delta f_1$, namely $\left|\widehat{\widetilde{G}}_2(f_2)\right|^2$ and $\left|\widehat{\widetilde{G}}_2(f_2)\right|^2\left[1-u_2^2(f_2)\right]$, respectively, see Eqs.~\eqref{eq:open_loop_observables_b} and~\eqref{eq:open_loop_observables_a}. For comparison, the modulus of the transfer function, $\left|\widehat{\widetilde{G}}_2(f_2)\right|$, is also reported. The first expression produces a resonance peak that is narrower than that of the transfer-function modulus, since it corresponds to its square. In the high-$Q$ limit, the full width at half maximum of $\left|\widehat{\widetilde{G}}_2(f_2)\right|^2$ is reduced by a factor $\sqrt{3}$ compared with $\left|\widehat{\widetilde{G}}_2(f_2)\right|$. The second expression changes sign at resonance. It displays a sharp positive lobe just below the resonance frequency and a negative lobe just above it, before gradually decaying on both sides of the peak, as predicted, see Sec.~\ref{sec:consequences}. Such behavior is not expected from the DC-modulated electrostatic contribution, or from conservative interaction-mediated effects, which act through the $\beta_1$ and $k_{\mathrm{int},1}^{(c)}$ terms in Eq.~\eqref{eq:open_loop_observables_a} and only tend to decrease the effective stiffness. If experimentally observed, such a signature can therefore only be assigned to the inverse heterodyne effect.

Thus, $V_d(f_2)$ and $\Delta f_1(f_2)$ were measured. The results are reported in Fig.~\ref{fig:fd_df1_vs_f2}(c) and Fig.~\ref{fig:fd_df1_vs_f2}(d), respectively. To do so, the $z$-feedback loop was kept open during the measurements. Because the amplitude of the inverse heterodyne effect on the first-eigenmode observables is small, $\alpha_1$ was made as large as possible. The experimental parameters were set to $V_{\mathrm{DC}}=+1$~V and $U_{\mathrm{mod}}=1$~V, and the initial $\Delta f_1$ value, set before opening the $z$-feedback loop, was fixed to $-100$~Hz in order to reduce $z_c$. The experimental curves of $V_d(f_2)$ and $\Delta f_1(f_2)$ closely agree with the predictions of the present formalism and provide a particularly discriminating test of the inverse heterodyne mechanism. The drive-voltage signal $V_d(f_2)$ exhibits a narrow resonant peak centered at the shifted second-eigenmode resonance, as expected from Eq.~\eqref{eq:open_loop_observables_b}, where the inverse heterodyne contribution to the first-eigenmode power balance scales as $\left|\widehat{\widetilde{G}}_2(f_2)\right|^2$. This squared-transfer-function dependence is a key signature of the effect: the experimentally determined full width at half maximum of $V_d(f_2)$ matches the theoretical linewidth and is approximately $1/\sqrt{3}$ times narrower than the linewidth of $\left|\widehat{\widetilde{G}}_2(f_2)\right|$. This narrowing confirms that the measured dissipative response is not simply proportional to the second-eigenmode amplitude, but to the energy-transfer channel mediated by the inverse heterodyne coupling. The simultaneously acquired $\Delta f_1(f_2)$ signal further supports this interpretation. Its weaker, sign-changing response around resonance follows the expected dependence on $\left|\widehat{\widetilde{G}}_2(f_2)\right|^2\left[1-u_2^2(f_2)\right]$, with opposite contributions below and above the resonance. When $f_2\lesssim\widetilde{f}_{2,0}$, the inverse heterodyne contribution increases the effective stiffness-like contribution entering the first-eigenmode frequency shift, thereby making $\Delta f_1$ more negative. Conversely, when $f_2\gtrsim\widetilde{f}_{2,0}$, this contribution changes sign and reduces the effective stiffness-like contribution, making $\Delta f_1$ less negative. This dispersive-like signature, corresponding to an apparent decrease or increase of the first-eigenmode effective stiffness depending on the side of the second-eigenmode resonance, cannot be accounted for by the DC-modulated electrostatic contribution, which does not produce a sign reversal around $f_2=\widetilde{f}_{2,0}$. Together, the resonant dissipation peak in $V_d(f_2)$ and the weaker dispersive response in $\Delta f_1(f_2)$ provide direct experimental evidence that the heterodyne-driven second eigenmode acts back on the first eigenmode, thereby confirming the occurrence of inverse heterodyne coupling and inter-mode energy exchange.

\subsubsection{Evolution of $F_d$ and $\Delta f_1$ with $V_{\mathrm{DC}}$}
\label{sec:fd_df1_vs_vdc}

\begin{figure}[htbp]
    \centering
    \includegraphics[width=\textwidth]{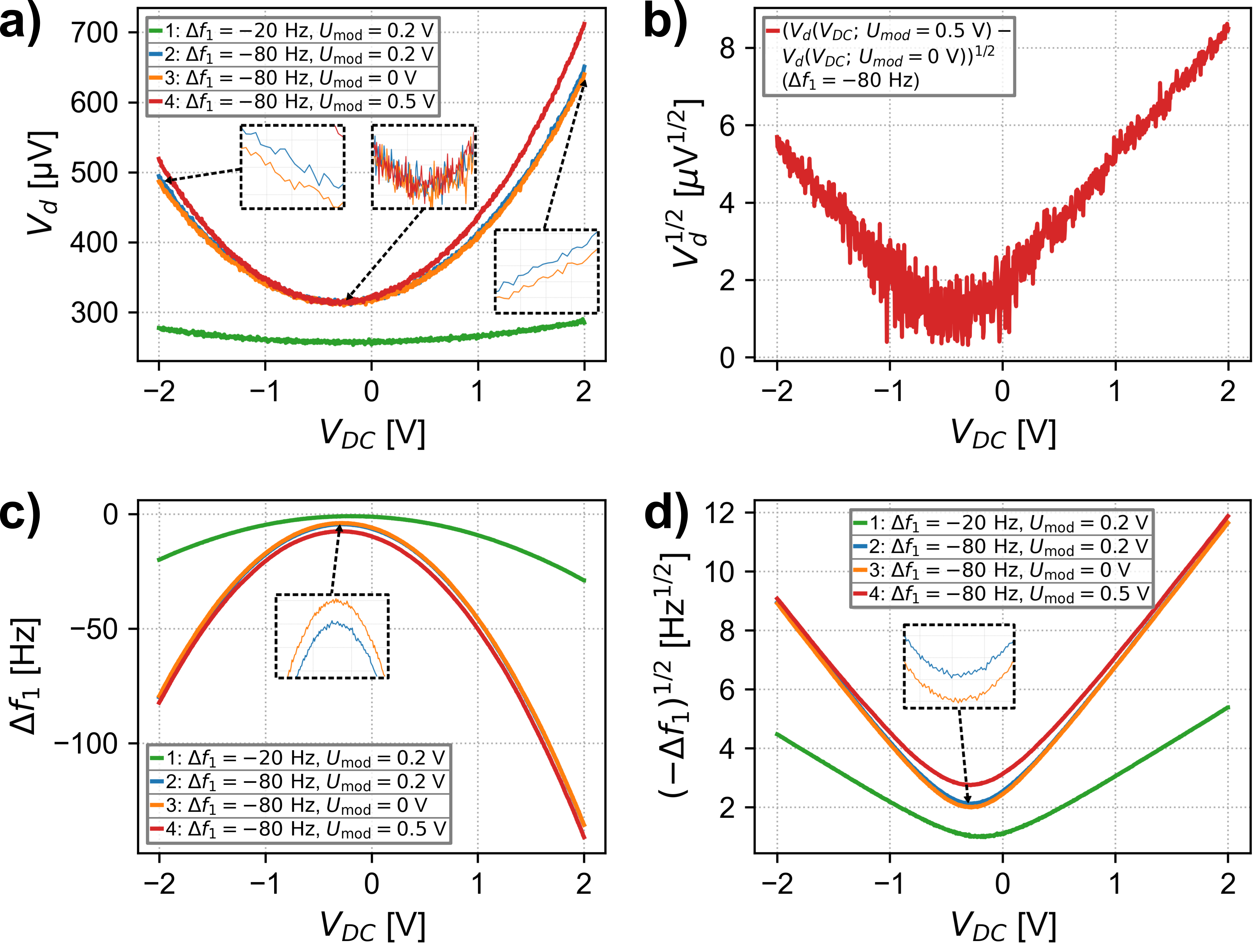}
    \caption{(a) $V_d(V_{\mathrm{DC}})$ acquired with the $z$-feedback loop open, for different values of $\Delta f_1$ and $U_{\mathrm{mod}}$. The qualitatively parabolic dependence is primarily attributed to a dissipative interaction encompassed in the term $k_{\mathrm{int},1}^{(d)}$. The inverse heterodyne effect appears as an additional, smaller contribution; see text. The curves exhibit a well-defined minimum at $V_{\mathrm{DC}}=V_{\mathrm{cpd}}$. (b) $\left[V_d(V_{\mathrm{DC}};U_{\mathrm{mod}}=0.5~\mathrm{V})-V_d(V_{\mathrm{DC}};U_{\mathrm{mod}}=0~\mathrm{V})\right]^{1/2}$ as a function of $V_{\mathrm{DC}}$ (linearized representation), which yields on both sides of the CPD, $V_{\mathrm{DC}}=V_{\mathrm{cpd}}$, a linear behavior, as predicted from the inverse heterodyne contribution to dissipation. (c) $\Delta f_1(V_{\mathrm{DC}})$ acquired simultaneously. The observed parabolic behavior is characteristic of conservative electrostatic interactions, with a maximum at $V_{\mathrm{DC}}=V_{\mathrm{cpd}}$. To a lesser extent, the curvature is also influenced by the inverse heterodyne effect; see text. (d) Linearized representation of $\Delta f_1(V_{\mathrm{DC}})$, obtained by plotting $\left(-\Delta f_1\right)^{1/2}(V_{\mathrm{DC}})$ for curves~1--4.}
    \label{fig:fd_df1_vs_vdc}
\end{figure}

The dependence of $F_d$ (experimentally measured through $V_d$) and $\Delta f_1$ on $V_{\mathrm{DC}}$ was measured under various conditions, see Table~\ref{tab:experimental_conditions_vdc}. For each condition, $V_{\mathrm{DC}}=-2$~V was initially applied to the tip. The $z$-feedback loop was then opened, and $V_{\mathrm{DC}}$ was swept up to $+2$~V. The results are reported in Fig.~\ref{fig:fd_df1_vs_vdc}.

Fig.~\ref{fig:fd_df1_vs_vdc}(a) shows that the dissipative effects qualitatively follow a parabolic dependence on $V_{\mathrm{DC}}$, with a minimum at $V_{\mathrm{DC}}=V_{\mathrm{cpd}}$. This behavior is predicted by Eq.~\eqref{eq:open_loop_observables_b} through the inverse heterodyne effect, since $\alpha_1\alpha_2\propto\left(V_{\mathrm{DC}}-V_{\mathrm{cpd}}\right)^2U_{\mathrm{mod}}^2$. Based on this equation, a qualitatively similar parabolic behavior should be observed when sweeping either $V_{\mathrm{DC}}$ or $U_{\mathrm{mod}}$. However, as discussed in Sec.~\ref{sec:fd_df1_vs_umod}, the parabolic behavior is more readily observed when sweeping $V_{\mathrm{DC}}$ than when sweeping $U_{\mathrm{mod}}$. This means that the parabolic $V_{\mathrm{DC}}$ dependence of $V_d$ is always observed without requiring specific conditions, such as minimized $z_c$, increased $z_{1,0}$, or increased $U_{\mathrm{mod}}$. In other words, the effect cannot be fully attributed to a CG effect. Furthermore, no parabolic dependence should occur when $U_{\mathrm{mod}}=0$, which is not what is observed, see curve~3 in Fig.~\ref{fig:fd_df1_vs_vdc}(a). The consistency between the model and the experimental results suggests that the inverse heterodyne effect is also at play here. However, as discussed later, its magnitude is smaller than that of another dissipative effect with a similar dependence on $V_{\mathrm{DC}}$. This latter effect is not explicitly accounted for in the present approach, but it can be interpreted as being included in the dissipative interaction-induced stiffness-like quantity $k_{\mathrm{int},1}^{(d)}$ in Eq.~\eqref{eq:open_loop_observables_b}. A plausible explanation for the hidden $V_{\mathrm{DC}}$ dependence of this term is discussed in Sec.~\ref{sec:discussion_generalization}.

Fig.~\ref{fig:fd_df1_vs_vdc}(c) shows the evolution of $\Delta f_1(V_{\mathrm{DC}})$. The curves exhibit a downward-opening parabolic dependence, with a maximum at $V_{\mathrm{DC}}=V_{\mathrm{cpd}}$. This well-known behavior is consistent with the occurrence of the term $\beta_1$ in Eq.~\eqref{eq:open_loop_observables_a}, which reflects the $\left(V_{\mathrm{DC}}-V_{\mathrm{cpd}}\right)^2$ dependence of the DC-modulated electrostatic force, see Eqs.~\eqref{eq:dc_modulated_electrostatic_forces_a} and~\eqref{eq:beta_coefficients}. The less intuitive result here is that part of the parabolic dependence of the frequency shift also stems from the inverse heterodyne effect, through the term $\alpha_1\alpha_2\propto\left(V_{\mathrm{DC}}-V_{\mathrm{cpd}}\right)^2$ in Eq.~\eqref{eq:open_loop_observables_a}. For clarity, Fig.~\ref{fig:fd_df1_vs_vdc}(d) shows the linearized representation obtained by plotting $\left(-\Delta f_1\right)^{1/2}(V_{\mathrm{DC}})$, which highlights the dependence of $\Delta f_1$ on $\left(V_{\mathrm{DC}}-V_{\mathrm{cpd}}\right)^2$. For all curves, the maximum of the parabola occurs at $V_{\mathrm{DC}}=V_{\mathrm{cpd}}$, where the DC-modulated electrostatic contribution is minimized since $\beta_1$ reaches its minimum, and the inverse heterodyne effect vanishes since $\alpha_1$ and $\alpha_2$ are both zero.

Comparative measurements of $V_d(V_{\mathrm{DC}})$ and $\Delta f_1(V_{\mathrm{DC}})$ are now performed for two distinct values of $z_c$, with $U_{\mathrm{mod}}=0.2$~V, see curves~1 and~2 in Fig.~\ref{fig:fd_df1_vs_vdc}(a) and Fig.~\ref{fig:fd_df1_vs_vdc}(c). All other parameters are kept unchanged. Before opening the $z$-feedback loop, $\Delta f_1=-20$~Hz was first set for curve~1, then $\Delta f_1=-80$~Hz for curve~2, thereby reducing $z_c$.

The comparison between curves~1 and~2 of $V_d(V_{\mathrm{DC}})$ shows that the curvature of the approximately parabolic dependence increases as $\Delta f_1$ decreases, i.e. as $z_c$ decreases. The vertical offset of the curve also increases when $z_c$ decreases, a behavior that is not described by the inverse heterodyne effect. Again, a plausible $z_c$ dependence of the term $k_{\mathrm{int},1}^{(d)}(z_c)$ of Eq.~\eqref{eq:open_loop_observables_b} is discussed in Sec.~\ref{sec:discussion_generalization}.

For the corresponding $\Delta f_1(V_{\mathrm{DC}})$ curves, the curvature of curve~1 is weaker than that of curve~2, consistent with the larger $z_c$. Based on Eq.~\eqref{eq:open_loop_observables_a}, this behavior is consistent with the inverse heterodyne effect, since the coefficients $K_i(z_c)$, where $i=1,2$, entering the expressions of $\alpha_i(z_c)$ and $\beta_i(z_c)$, see Eqs.~\eqref{eq:alpha_coefficients} and~\eqref{eq:beta_coefficients}, are expected to scale as $z_c^{-k}$, where $k>1$ depending on the capacitance model~\cite{Crowley2008a,Sugawara2012a}. As a result, depending on the CG regime, the $V_{\mathrm{DC}}$ dependence of both the DC-modulated electrostatic contribution and the inverse heterodyne contribution is modified. Curves~1 and~2 do not merge at $V_{\mathrm{DC}}=V_{\mathrm{cpd}}$ because of both the difference in conservative interaction, $k_{\mathrm{int},1}^{(c)}(z_c)$, and the fact that the DC-modulated electrostatic term $\beta_1(z_c)$ is $z_c$ dependent and does not vanish when $U_{\mathrm{mod}}\neq0$. This behavior was confirmed by measuring the dependence of $V_d$ and $\Delta f_1$ on $z_c$ while keeping the other parameters unchanged, see Sec.~SI-IV~B.

Lastly, measurements of $V_d(V_{\mathrm{DC}})$ and $\Delta f_1(V_{\mathrm{DC}})$ are compared for three distinct values of $U_{\mathrm{mod}}$, while keeping $z_c$ constant, see curves~2-4 in Fig.~\ref{fig:fd_df1_vs_vdc}(a) and Fig.~\ref{fig:fd_df1_vs_vdc}(c). Curves~3, 2, and~4 correspond to $U_{\mathrm{mod}}$ values of 0, 0.2, and 0.5~V, respectively. The value of $U_{\mathrm{mod}}$ was adjusted only after opening the $z$-feedback loop, in order to maintain comparable values of $z_c$ for the different measurements.

The $V_d(V_{\mathrm{DC}})$ curves exhibit a qualitatively parabolic curvature that increases quadratically with $U_{\mathrm{mod}}$, while the vertical offset of the curve remains unchanged. To demonstrate this, the curve acquired at $U_{\mathrm{mod}}=0$~V (curve~3) was subtracted from that acquired at $U_{\mathrm{mod}}=0.5$~V (curve~4), and the square root of the resulting signal was plotted as a function of $V_{\mathrm{DC}}$, as shown in Fig.~\ref{fig:fd_df1_vs_vdc}(b). In other words, the following quantity was plotted:

\begin{equation}\label{eq:Vd_VDC_difference}
\left[
V_d(V_{\mathrm{DC}};U_{\mathrm{mod}}=0.5\,\mathrm{V})
-
V_d(V_{\mathrm{DC}};U_{\mathrm{mod}}=0\,\mathrm{V})
\right]^{1/2}.
\end{equation}

This behavior is consistent with the expected inverse heterodyne contribution, which scales as $\alpha_1\alpha_2\propto\left(V_{\mathrm{DC}}-V_{\mathrm{cpd}}\right)^2U_{\mathrm{mod}}^2$. It also highlights the distinct responses of $V_d$ to $V_{\mathrm{DC}}$ and $U_{\mathrm{mod}}$: $U_{\mathrm{mod}}$ does not influence, or only negligibly influences, $V_d$ when $V_{\mathrm{DC}}=V_{\mathrm{cpd}}$, whereas $V_{\mathrm{DC}}$ affects $V_d$ even when $U_{\mathrm{mod}}=0$. This is not predicted by the inverse heterodyne effect, confirming the presence of an additional $V_{\mathrm{DC}}$-dependent dissipative contribution for curves~1-4, which does not scale, or scales only negligibly, with $U_{\mathrm{mod}}$. This interpretation is further supported by the measurements reported in Sec.~SI-IV~C, where the corresponding curves are recorded for different values of $V_{\mathrm{DC}}$. In contrast to Fig.~\ref{fig:fd_df1_vs_vdc}(b), curves~1-4 cannot be directly linearized by plotting $V_d^{1/2}$. Instead, they are dominated by the additional dissipative mechanism rather than by the inverse heterodyne effect: they do not follow a strictly parabolic law and therefore do not scale as $\left(V_{\mathrm{DC}}-V_{\mathrm{cpd}}\right)^2$, even though their overall shape may qualitatively suggest such a dependence.

The corresponding $\Delta f_1(V_{\mathrm{DC}})$ curves show that, at
$V_{\mathrm{DC}}=V_{\mathrm{cpd}}$, one has
$\left|\Delta f_1^{(3)}(V_{\mathrm{cpd}})\right|
<
\left|\Delta f_1^{(2)}(V_{\mathrm{cpd}})\right|
<
\left|\Delta f_1^{(4)}(V_{\mathrm{cpd}})\right|$. Moreover, each of these values follows a quadratic dependence on $U_{\mathrm{mod}}$. This confirms the DC-modulated electrostatic origin of the frequency shift through the coefficient $\beta_1$. The parabolic curvature also decreases slightly from curves~4 to~2 to~3, reflecting the contribution of the inverse heterodyne effect. For this effect, the parabolic coefficient of $\Delta f_1$ as a function of $V_{\mathrm{DC}}$ depends quadratically on $U_{\mathrm{mod}}$. However, both the magnitude and the sign of this curvature depend on the position of $f_2$ with respect to $\widetilde{f}_{2,0}$, as shown in the previous section. Since $\widetilde{f}_{2,0}$ is not tracked, $f_2$ is slightly shifted from resonance. In the present case, it lies on the right-hand side of the resonance peak, where the conservative component of the inverse heterodyne effect increases the effective stiffness of the system. This explains why the curvature of the parabolas decreases as $U_{\mathrm{mod}}$ increases, due to the inverse heterodyne effect.

Complementary measurements of the $V_d(V_{\mathrm{DC}})$ and $\Delta f_1(V_{\mathrm{DC}})$ curves are discussed in Sec.~SI-IV~C.

\subsubsection{Evolution of $F_d$ and $\Delta f_1$ with $U_{\mathrm{mod}}$}
\label{sec:fd_df1_vs_umod}

\begin{figure}[htbp]
    \centering
    \includegraphics[width=\textwidth]{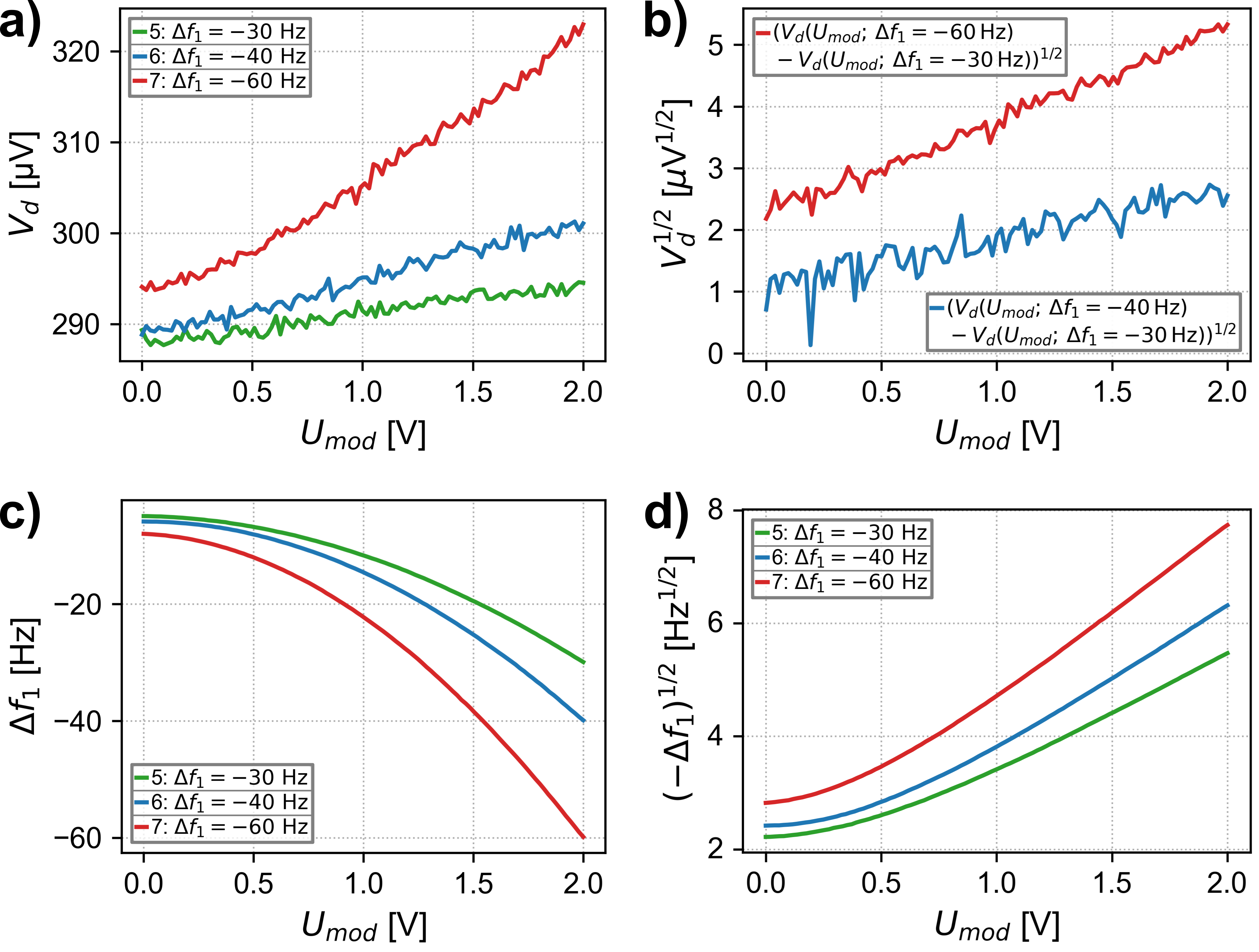}
    \caption{(a) $V_d(U_{\mathrm{mod}})$ acquired with the $z$-feedback loop open for different values of $\Delta f_1$, and therefore for different values of the average tip--surface distance $z_c$. The quadratic dependence is mainly attributed to the inverse heterodyne effect, which scales as $\alpha_1\alpha_2\propto U_{\mathrm{mod}}^2$. The curves exhibit a well-defined minimum at $U_{\mathrm{mod}}=0$. The fact that the curves do not merge when $U_{\mathrm{mod}}=0$ is attributed to an additional dissipative interaction encompassed in the term $k_{\mathrm{int},1}^{(d)}(z_c)$, whose magnitude depends on the average tip--surface distance $z_c$; see text. (b) Linearized representation of the $U_{\mathrm{mod}}$-dependent dissipative contrast between setpoints, obtained by plotting $\left[V_d(U_{\mathrm{mod}};\Delta f_1=-60~\mathrm{Hz})-V_d(U_{\mathrm{mod}};\Delta f_1=-30~\mathrm{Hz})\right]^{1/2}$ (red) and $\left[V_d(U_{\mathrm{mod}};\Delta f_1=-40~\mathrm{Hz})-V_d(U_{\mathrm{mod}};\Delta f_1=-30~\mathrm{Hz})\right]^{1/2}$ (blue) as functions of $U_{\mathrm{mod}}$. This representation highlights the inverse heterodyne contribution through the approximately linear dependence on $U_{\mathrm{mod}}$, consistent with $V_d(U_{\mathrm{mod}})\propto U_{\mathrm{mod}}^2$. (c) $\Delta f_1(U_{\mathrm{mod}})$ acquired simultaneously. The observed quadratic behavior is characteristic of conservative electrostatic interactions, with a maximum when $U_{\mathrm{mod}}=0$. To a lesser extent, the curvature is also influenced by the inverse heterodyne effect; see text. (d) Linearized representation of $\Delta f_1(U_{\mathrm{mod}})$, obtained by plotting $\left(-\Delta f_1\right)^{1/2}(U_{\mathrm{mod}})$ for curves~5--7.}
    \label{fig:fd_df1_vs_umod}
\end{figure}

$V_d(U_{\mathrm{mod}})$ and $\Delta f_1(U_{\mathrm{mod}})$ are now investigated under several conditions, see measurements~5, 6, and~7 in Table~\ref{tab:experimental_conditions_vdc}. For all measurements, $U_{\mathrm{mod}}$ was initially set to $+2$~V. The $z$-feedback loop was then opened so that $z_c$ remained constant throughout the voltage sweep down to $U_{\mathrm{mod}}=0$~V. Fig.~\ref{fig:fd_df1_vs_umod} shows the evolution of $V_d$ and $\Delta f_1$ with $U_{\mathrm{mod}}$.

Overall, the $V_d(U_{\mathrm{mod}})$ curves exhibit a half-parabolic shape, with a minimum at $U_{\mathrm{mod}}=0$~V, as expected for a quadratic contribution. These observations are consistent with Eq.~\eqref{eq:open_loop_observables_b}, which states that $V_d$ scales, through the inverse heterodyne effect, as $\alpha_1\alpha_2\propto U_{\mathrm{mod}}^2$. When $U_{\mathrm{mod}}=0$, the inverse effect vanishes and the $V_d(U_{\mathrm{mod}})$ curve reaches its minimum. Following the same approach as in the previous section, this quadratic dependence is further validated by plotting $V_d^{1/2}(U_{\mathrm{mod}})$ for the differences between curves~7 and~5 and between curves~6 and~5, see Fig.~\ref{fig:fd_df1_vs_umod}(b). In other words, the following quantities were plotted:

\begin{subequations}\label{eq:Vd_Umod_difference}
\begin{empheq}[left=\empheqlbrace]{align}
\left[
V_d(U_{\mathrm{mod}};\Delta f_1=-60\,\mathrm{Hz})
-
V_d(U_{\mathrm{mod}};\Delta f_1=-30\,\mathrm{Hz})
\right]^{1/2},
\label{eq:Vd_Umod_difference_a}
\\
\left[
V_d(U_{\mathrm{mod}};\Delta f_1=-40\,\mathrm{Hz})
-
V_d(U_{\mathrm{mod}};\Delta f_1=-30\,\mathrm{Hz})
\right]^{1/2}.
\label{eq:Vd_Umod_difference_b}
\end{empheq}
\end{subequations}

The resulting linear trends confirm that $V_d$ scales with $U_{\mathrm{mod}}^2$.

The corresponding $\Delta f_1(U_{\mathrm{mod}})$ curves exhibit downward-opening half-parabolic shapes, with a maximum at $U_{\mathrm{mod}}=0$, consistent with the behavior expected from Eq.~\eqref{eq:open_loop_observables_a}. This behavior stems from the combined contributions of the DC-modulated electrostatic interaction, described by the term $\beta_1$, see Eq.~\eqref{eq:beta_coefficients}, and of the inverse heterodyne effect, through the term $\alpha_1\alpha_2\propto U_{\mathrm{mod}}^2$. For clarity, Fig.~\ref{fig:fd_df1_vs_umod}(d) shows the linearized representation obtained by plotting $\left(-\Delta f_1\right)^{1/2}(U_{\mathrm{mod}})$, which highlights the $U_{\mathrm{mod}}^2$ dependence for curves~5-7. Moreover, the term $\beta_1$ reaches its minimum and the inverse heterodyne effect vanishes when $U_{\mathrm{mod}}=0$.

Curves~5, 6, and~7 reported in Fig.~\ref{fig:fd_df1_vs_umod}(a) were acquired for progressively smaller values of $z_c$, by setting $\Delta f_1$ to $-30$, $-40$, and $-60$~Hz, respectively, prior to opening the $z$-feedback loop. From curves~5 to~7, it can be seen that the curvature of the $V_d(U_{\mathrm{mod}})$ curves increases as $z_c$ decreases. This behavior is consistent with the inverse heterodyne effect, since the coefficients $K_i(z_c)$, where $i=1,2$, entering the expressions of the coefficients $\alpha_i(z_c)$, see Eq.~\eqref{eq:alpha_coefficients}, are expected to scale as $z_c^{-k}$, where $k>1$ depending on the capacitance model. It can also be noticed that the curvature of the $\Delta f_1(U_{\mathrm{mod}})$ curves increases as $z_c$ decreases. This trend is consistent with the strengthening of both the DC-modulated electrostatic and inverse heterodyne contributions, following arguments similar to those discussed above, through the $\beta_1(z_c)$ term and the product $\alpha_1(z_c)\alpha_2(z_c)\propto K_1(z_c)K_2(z_c)\propto z_c^{-2k}$. It should be recalled that the complete evolution of $V_d$ and $\Delta f_1$ with $z_c$ is reported in the SI, see Sec.~SI-IV~B.

In Fig.~\ref{fig:fd_df1_vs_umod}(a), the $V_d(U_{\mathrm{mod}})$ curves do not merge when $U_{\mathrm{mod}}=0$ because of the additional dissipative effect previously mentioned, see Sec.~\ref{sec:fd_df1_vs_vdc}, and discussed in Sec.~\ref{sec:discussion_generalization}. As discussed above, this contribution varies quadratically with $V_{\mathrm{DC}}-V_{\mathrm{cpd}}$, independently of, or only weakly dependent on, $U_{\mathrm{mod}}$, but it depends on $z_c$. Consequently, the strength of this contribution is not identical for curves~5, 6, and~7, although it does not significantly affect the curvature of the $V_d(U_{\mathrm{mod}})$ curves. The $\Delta f_1(U_{\mathrm{mod}})$ curves do not merge either when $U_{\mathrm{mod}}=0$, because the DC-modulated electrostatic contribution and the interaction-induced shift, described by the $\beta_1(z_c)$ and $k_{\mathrm{int},1}^{(c)}(z_c)$ terms, respectively, are non-zero and vary with $z_c$.

Complementary measurements of the $V_d(U_{\mathrm{mod}})$ and $\Delta f_1(U_{\mathrm{mod}})$ curves are discussed in Sec.~SI-IV~C.

\clearpage

\section{Discussion and generalization}
\label{sec:discussion_generalization}

The analytical model developed here was validated in the previous section and through the experimental measurements presented in Sec.~SI-IV by examining the dependence of multiple observables on key measurement parameters in nc-AFM open-loop AM-He-KPFM. Most of the theoretical predictions were observed experimentally, enabling detailed interpretations of the measurements to be provided. The direct heterodyne effect was investigated through the evolution of $z_{2,0}$ as a function of the measurement parameters and was clearly identified. The influence of the inverse heterodyne effect on the first-eigenmode observables, $F_d$ and $\Delta f_1$, was also clearly revealed, notably by sweeping $f_2$, $V_{\mathrm{DC}}$, and $U_{\mathrm{mod}}$. This indicates that electromechanically induced energy transfer occurs between eigenmodes. Lastly, the observations show that the inverse heterodyne effect has a stronger influence on $F_d$ than on $\Delta f_1$. Indeed, for measurements performed close to resonance, the dissipative component of the interaction related to the inverse heterodyne effect largely prevails over its conservative component, owing to the contribution of the second-eigenmode transfer-function term, see Sec.~\ref{sec:fd_df1_vs_f2}.

The present results should also be distinguished from earlier dissipative KPFM approaches, in which dissipation was intentionally used as the primary electrostatic measurement channel~\cite{Miyahara2015a,Miyahara2017a,Sadewasser2018a}. In those schemes, the central point was that electrostatic excitation could be efficiently detected through the cantilever dissipation signal. By contrast, it is shown here that, in open-loop AM-He-KPFM, dissipation may also arise as the signature of an inverse heterodyne back-action between eigenmodes. In this sense, the dissipation channel is not only a convenient observable, but also the manifestation of an intrinsic inter-mode energy-transfer process induced by heterodyne frequency conversion. This distinction is important because it changes the physical interpretation of dissipation-based contrast in open-loop AM-He-KPFM and, more generally, in multimode force-microscopy schemes involving frequency conversion.

A more detailed experimental analysis of $F_d$ and $\Delta f_1$ as functions of $V_{\mathrm{DC}}$ and $U_{\mathrm{mod}}$ revealed that, besides the inverse heterodyne effect, an additional dissipative contribution exists, which scales with both $V_{\mathrm{DC}}$ and $z_c$. To make this observation consistent with the present formalism, this effect can be formally absorbed into the term $k_{\mathrm{int},1}^{(d)}$, see Eq.~\eqref{eq:open_loop_observables_b}. Such a $V_{\mathrm{DC}}$ dependence of the dissipation channel in nc-AFM was reported early on for semiconducting and metallic substrates~\cite{Denk1991,Stowe1999,Dorofeyev1999,loppacher00b,pfeiffer04a,Dwyer2019a,Hasan2022a}. From a theoretical standpoint, this effect was generally interpreted as non-contact friction arising from fluctuating electric forces above the sample surface, generated by molecular and charge-carrier dynamics, inducing cantilever frequency fluctuations and ultimately measurable Joule heat dissipation in the system~\cite{Dorofeyev1999,Lekkala2013a,Loring2022a}. In both samples investigated here, namely the silicon sample discussed in the main text and an Au(111) sample not shown, similar observations were made: the additional dissipative process is more pronounced than the inverse heterodyne effect. In fact, the impact of the inverse heterodyne effect on first-eigenmode dissipation is more clearly isolated in experiments carried out in DHe-KPFM. In this configuration, the inverse heterodyne contribution is observed without the additional $V_{\mathrm{DC}}$-dependent dissipative effect identified in open-loop AM-He-KPFM. In the SI, a straightforward generalization of the present formalism to DHe-KPFM is described. In that case, for all tested values of $V_{\mathrm{DC}}$, $F_d$ exhibits a quadratic dependence as a function of the intermodulation products. Complementary measurements carried out under electrical pumping on a graphene sample on silicon carbide are consistent with this prediction, see Sec.~SI-V.

This work focuses on describing the behavior of the observables during nc-AFM experiments carried out in open-loop AM-He-KPFM. In the ideal closed-loop KPFM limit, the active DC potential-compensation loop cancels the heterodyne force component by minimizing the residual bias term $V_{\mathrm{DC}}-V_{\mathrm{cpd}}$. As a result, $\alpha_i$ ($i=1,2$) and the second-eigenmode amplitude $z_{2,0}$ are minimized, and the inverse heterodyne contribution vanishes to first order. The remaining parameters and observables, namely the first-eigenmode amplitude, driving force, and frequency shift, then reduce approximately to those derived from a simpler monomodal formalism, see Eq.~\eqref{eq:open_loop_observables} in the limiting case $\alpha_i\rightarrow0$ for $i=1,2$, and $z_{2,0}\rightarrow0$. Thus, the present approach might seem irrelevant when discussing ideal closed-loop KPFM. However, this conclusion would be misleading, since the present formalism also describes the residual bimodal dynamics that may exist before or during compensation when the heterodyne force is not perfectly canceled. This operating condition should therefore be optimized, since the efficiency of the heterodyne drive mechanism was shown to depend on the choice of the second-eigenmode frequency. This seemingly trivial point appears not to have been discussed in detail in the literature. In practice, it is common to set the second-eigenmode frequency, $f_2$, to the same value as the free resonance frequency, $f_{2,0}$. However, it has been demonstrated that, if the second eigenmode undergoes a significant frequency shift, the efficiency of the direct heterodyne drive can be strongly affected. The impact on CPD measurement sensitivity, in terms of bandwidth or signal-to-noise ratio, remains to be quantified precisely, but should not be neglected. A well-known but often overlooked fact is that the AC bias modulation used in KPFM induces a DC attractive force, and consequently a frequency shift, that cannot be canceled, contrary to the CPD term. This force, proportional to the squared modulation depth, shifts the second-eigenmode frequency even in closed-loop KPFM. Thus, the analytical derivation presented here highlights that, to fully benefit from resonance-enhanced sensitivity, the modulation depth should be kept reasonably small.

Lastly, sideband coupling and frequency conversion are not unique to KPFM. They have also long been used in PiFM. PiFM detects an effective photo-induced dipole-dipole force, often referred to as the optical force, which is related to the optical polarizability of the sample. Like KPFM, it can be implemented in an AM heterodyne variant~\cite{Yamanishi2017a,Shcherbakov2025a}. In AM-He-PiFM, sidebands arise from the product of the cantilever deflection and the modulated optical force, with the optical modulation frequency set equal to the difference between the cantilever eigenmode frequencies. The first eigenmode is used for topographic imaging, while the second one is used to demodulate the optical force. The experimental methodology is therefore identical to that of open-loop AM-He-KPFM. The analytical treatment developed here can be directly transposed to AM-He-PiFM. PiFM observables have previously been analyzed within the framework of multimodal AFM~\cite{Jahng2016a}, but the present study provides additional insight into the role of heterodyne back-action and inter-mode energy exchange.

More broadly, the present results place open-loop AM-He-KPFM within a wider class of driven multimode systems in which nonlinear coupling gives rise to inter-mode energy transfer and measurable back-action. Similar phenomena have been reported in internally resonant micromechanical oscillators, where coherent energy exchange between vibrational modes can sustain or restore the motion of a principal mode, and more generally in micro- and nanomechanical resonators displaying intermodal coupling, internal resonance, and synchronization. Related energy-transfer processes have also been demonstrated in parametrically coupled mechanical modes. In this sense, the inverse heterodyne effect identified here is not only relevant to KPFM instrumentation, but also connects heterodyne force microscopy to a broader dynamical framework in which dissipation can reveal mode-to-mode energy exchange induced by frequency conversion~\cite{Chen2017a,Asadi2018a,Fu2019a}.

\clearpage

\section{Conclusion}
\label{sec:conclusion}

In this work, it is established that open-loop AM-He-KPFM intrinsically supports inverse heterodyne coupling between cantilever eigenmodes. Beyond the well-known direct heterodyne actuation of the second eigenmode, heterodyne frequency conversion generates a back-action force component that acts at the fundamental eigenmode frequency and couples the conservative and dissipative dynamics of the two modes. Incorporating these force components into the bimodal virial and power-balance formalism yields analytical expressions for the key observables of nc-AFM open-loop AM-He-KPFM: the frequency shift and dissipated energy of the first eigenmode, and the amplitude and phase of the second eigenmode. A central outcome is that inverse heterodyne coupling is detected most sensitively through energy dissipation, because the relevant contribution scales with the squared transfer function of the second eigenmode and therefore produces a narrow resonant signature when sweeping the demodulation frequency.

Crucially, the present study is built on a rigorous, non-truncated treatment of the CG dynamics under bimodal motion, developed in the companion manuscript~\cite{ValloireSubmittedCG}. The companion manuscript provides the complete derivation, truncation regimes, and numerical validation of the effective CG coefficients that enter the heterodyne force components and ultimately control the strength and distance dependence of inverse heterodyne back-action. This connection is not merely ancillary: it establishes the formal consistency of the heterodyne force decomposition used here and defines the regime of validity of the analytical expressions derived for the observables.

The UHV measurements provide direct evidence of inverse heterodyne back-action. In particular, sweeping the demodulation frequency around the second-eigenmode resonance isolates a sharply resonant dissipation feature with the predicted linewidth and symmetry, while the corresponding effect on the frequency shift remains weaker. Voltage-dependent measurements further confirm the characteristic quadratic scaling expected for heterodyne-induced back-action. Together, these results demonstrate that inverse heterodyne effects are an intrinsic ingredient of open-loop heterodyne force microscopy and must be accounted for when interpreting dissipation-based contrast and quantitative electrostatic measurements.

More broadly, this paper and its companion theoretical counterpart establish a coherent and practical framework for heterodyne electrostatic force microscopy. This framework is directly relevant to advanced heterodyne implementations, such as dual-heterodyne KPFM and heterodyne photo-induced force microscopy, and to a wider class of driven multimode resonators relying on frequency conversion, where similar back-action mechanisms are expected.

\clearpage

\begin{acknowledgments}
This work was supported by the Centre National de la Recherche Scientifique (CNRS) and Aix-Marseille Université. The authors thank the ANR funding agency for financial support of the PESOS project
(ANR-23-CE09-0038, H.V., S.C., C.L., L.N. and B.G.) and the SuperZIC project (ANR-22-CE09-0020, S.C., C.L. and L.N.). The authors thank O.~Pilone and F.~Para (IM2NP) for experimental support, and K.~Courouble (ST Microelectronics) for providing the passivated c-Si sample.
\end{acknowledgments}

\section*{Author Contributions}

H.V.: conceptualization; methodology; software; data acquisition and curation; formal analysis; investigation; validation; writing--original draft preparation; writing--review and editing. S.C.: investigation; validation; writing--review and editing. C.L.: investigation; validation; writing--review and editing. L.N.: conceptualization; methodology; data acquisition and curation; formal analysis; investigation; validation; writing--original draft preparation; writing--review and editing; funding acquisition; supervision. B.G.: conceptualization; methodology; data acquisition and curation; formal analysis; investigation; validation; writing--original draft preparation; writing--review and editing; funding acquisition; supervision; project administration.

\section*{Data Availability Statement}

The data that support the findings of this study are available from the corresponding author upon reasonable request.

\clearpage

\makeatletter
\renewcommand{\bibsection}{%
  \section*{References}%
}
\makeatother

\bibliographystyle{apsrev4-2}
\bibliography{uhvafm_bib}

\clearpage

\section*{Figures}\label{sec_figures}

\clearpage

\section*{Tables}\label{sec_tables}

\begin{table}[htbp]
\centering
\caption{Experimental parameters.}
\label{tab:cantilever_parameters}
\renewcommand{\arraystretch}{1.2}
\begin{adjustbox}{width=\textwidth}
\begin{tabular}{|c|c|c|}
\hline
\textbf{Cantilever:} & \textbf{Eigenmode 1} & \textbf{Eigenmode 2} \\
\hline
Eigenfrequency
& $f_{1,0}=161.225~\mathrm{kHz}$
& $f_{2,0}=999.425~\mathrm{kHz}\approx 6.2\,f_{1,0}$ \\
\hline
Super-frequency
& \multicolumn{2}{c|}{$f_s \approx 25~\mathrm{Hz}$} \\
\hline
Super-period
& \multicolumn{2}{c|}{$T_s = 1/f_s \approx 40~\mathrm{ms}$} \\
\hline
Stiffness (nominal)
& $k_1 = 48~\mathrm{N/m}$
& $k_2 = 39.3\,k_1 = 1886~\mathrm{N/m}$ \\
\hline
Q-factor
& $Q_1 = 20680$
& $Q_2 = 2200$ \\
\hline
Oscillation amplitude
& $z_{1,0}=(12\pm0.5)~\mathrm{nm}$
& $z_{2,0}=z_{1,0}/68=(0.18\pm0.02)~\mathrm{nm}$ \\
\hline
Phase
& $\Phi_1=-\pi/2$
& $\Phi_2=\mathrm{varied}$ \\
\hline
\multicolumn{3}{|c|}{\textbf{Electrostatic modulation:}} \\
\hline
Modulation frequency
& \multicolumn{2}{c|}{$f_{\mathrm{mod}}=f_{2,0}-f_{1,0}=838.2~\mathrm{kHz}$} \\
\hline
Super-super-frequency
& \multicolumn{2}{c|}{$f_{ss}=f_s\approx 25~\mathrm{Hz}$} \\
\hline
Super-super-period
& \multicolumn{2}{c|}{$T_{ss}=T_s\approx 40~\mathrm{ms}$} \\
\hline
DC bias
& \multicolumn{2}{c|}{$V_{\mathrm{DC}}=+200~\mathrm{mV}$} \\
\hline
CPD
& \multicolumn{2}{c|}{$V_{\mathrm{cpd}}=-200~\mathrm{mV}$} \\
\hline
Modulation depth
& \multicolumn{2}{c|}{$U_{\mathrm{mod}}=200~\mathrm{mV}$} \\
\hline
\end{tabular}
\end{adjustbox}
\end{table}

\begin{table}[htbp]
\centering
\caption{Experimental acquisition conditions.}
\label{tab:experimental_conditions_vdc}
\renewcommand{\arraystretch}{1.2}
\begin{adjustbox}{width=\textwidth}
\begin{tabular}{|c|c|c|c|c|c|c|}
\hline
Measurement \#
& $U_{\mathrm{mod}}$ [V]
& $V_{\mathrm{DC}}$ [V]
& $z_{1,0}$ [nm]
& $\Delta f_1$ [Hz]
& $f_2$ condition
& Swept parameter \\
\hline
1 & 0.2 & swept & 12 & $-20$
& \makecell{On resonance \\ (no tracking)}
& \makecell{$V_{\mathrm{DC}}$ swept from $-2$ to $+2$~V \\ in open $z$-feedback loop} \\
\hline
2 & 0.2 & swept & 12 & $-80$
& \makecell{On resonance \\ (no tracking)}
& \makecell{$V_{\mathrm{DC}}$ swept from $-2$ to $+2$~V \\ in open $z$-feedback loop} \\
\hline
3 & 0 & swept & 12 & $-80$
& \makecell{On resonance \\ (no tracking)}
& \makecell{$V_{\mathrm{DC}}$ swept from $-2$ to $+2$~V \\ in open $z$-feedback loop} \\
\hline
4 & 0.5 & swept & 12 & $-80$
& \makecell{On resonance \\ (no tracking)}
& \makecell{$V_{\mathrm{DC}}$ swept from $-2$ to $+2$~V \\ in open $z$-feedback loop} \\
\hline
5 & swept & 0.2 & 12 & $-30$
& \makecell{On resonance \\ (no tracking)}
& \makecell{$U_{\mathrm{mod}}$ swept from $2$ to $0$~V \\ in open $z$-feedback loop} \\
\hline
6 & swept & 0.2 & 12 & $-40$
& \makecell{On resonance \\ (no tracking)}
& \makecell{$U_{\mathrm{mod}}$ swept from $2$ to $0$~V \\ in open $z$-feedback loop} \\
\hline
7 & swept & 0.2 & 12 & $-60$
& \makecell{On resonance \\ (no tracking)}
& \makecell{$U_{\mathrm{mod}}$ swept from $2$ to $0$~V \\ in open $z$-feedback loop} \\
\hline
\end{tabular}
\end{adjustbox}
\end{table}

\clearpage

\end{document}